\documentclass[12pt,preprint]{aastex}
\usepackage[intlimits]{amsmath}
\usepackage{times}

\newcommand{\half}{\frac{1}{2}}

\shorttitle{Rectangular Masks}
\shortauthors{Vanderbei et al.}

\begin{document}


\title{Rectangular-Mask Coronagraphs for High-Contrast Imaging}


\author{Robert J. Vanderbei}
\affil{Operations Research and Financial Engineering, 
Princeton University}
\email{rvdb@princeton.edu}

\author{N. Jeremy Kasdin}
\affil{Mechanical and Aerospace Engineering, 
Princeton University}
\email{jkasdin@princeton.edu}

\author{David N. Spergel}
\affil{Astrophysics, 
Princeton University}
\email{dns@princeton.edu}


%
%


\begin{abstract}
We present yet another new family of masks for high-contrast imaging
as required for the to-be-built terrestrial planet finder space telescope.  
The ``best'' design
involves a square entrance pupil having a 4-vane spider, a square
image-plane mask containing a plus-sign shaped occulter to block the starlight
inside $0.6 \lambda/D$, and a Lyot-plane mask consisting of a rectangular
array of rectangular opennings.  Using Fraunhofer analysis, we show that 
the optical system can image a planet
$10^{-10}$ times as bright as an on-axis star in four rectangular
regions given by $\{ (\xi,\zeta): 1.4 < | \xi | < 20, 1.4 < | \zeta | < 20 \}$.

Since the design involves an image plane mask, pointing error is an issue.
We show that the design can tolerate pointing errors of about $0.05
\lambda/D$.

The inclusion of a 4-vane spider in the entrance pupil provides the
possibility to build a mirror-only on-axis system thereby
greatly reducing the negative effects of polarization.

Each of the masks can be realized as two masks consisting of stripes
of opaque material with the stripes oriented at right angles to each other.  
We call these striped masks {\em barcode} masks.  We show that it is
sufficient for the barcode masks by themselves to provide $10^{-5}$
contrast.  This then guarantees that the full system will provide the
required $10^{-10}$ contrast.

The rectangular Lyot-plane mask can be viewed as a binary-mask extension of 
Nisenson's apodized square aperture concept.
\end{abstract}




\section{Introduction}
\label{sec:intro}
Following our previous work on pupil-plane masks for high-contrast imaging
in the context of terrestrial planet finding
(see \cite{KVSL02,VSK02,VSK03}),
we present in this paper further new pupil masks as well as some
pupil-plane/image-plane combinations of masks.
Most of our earlier pupil-plane mask designs have an inner working angle
({\em IWA}) 
of $4 \lambda/D$, which implies a main mirror in the $12$ m range.
One of the main objectives leading to the designs presented in this 
paper was to significantly decrease this inner working angle.
If the inner working angle can be reduced to $2 \lambda/D$, then 
the main mirror can be dropped down to the $6$ m range and still 
be capable of studying the same set of stars.  This is our goal.

The simplest, but not best, pupil mask presented here
achieves a contrast ratio of $10^{-10}$ in four rectangular
regions given by $\{ (\xi,\zeta): 2 < | \xi | < 20, 2 < | \zeta | < 20 \}$.
The points of high-contrast that are closest to the center of the star's image
occur along the diagonals and therefore correspond 
to an inner working angle 
of $2\sqrt{2} = 2.8 \lambda/D$ (as
usual, $\lambda$ is wavelength and $D$ is aperture).  

\begin{figure}
\begin{center} 
\includegraphics[width=3.0in]{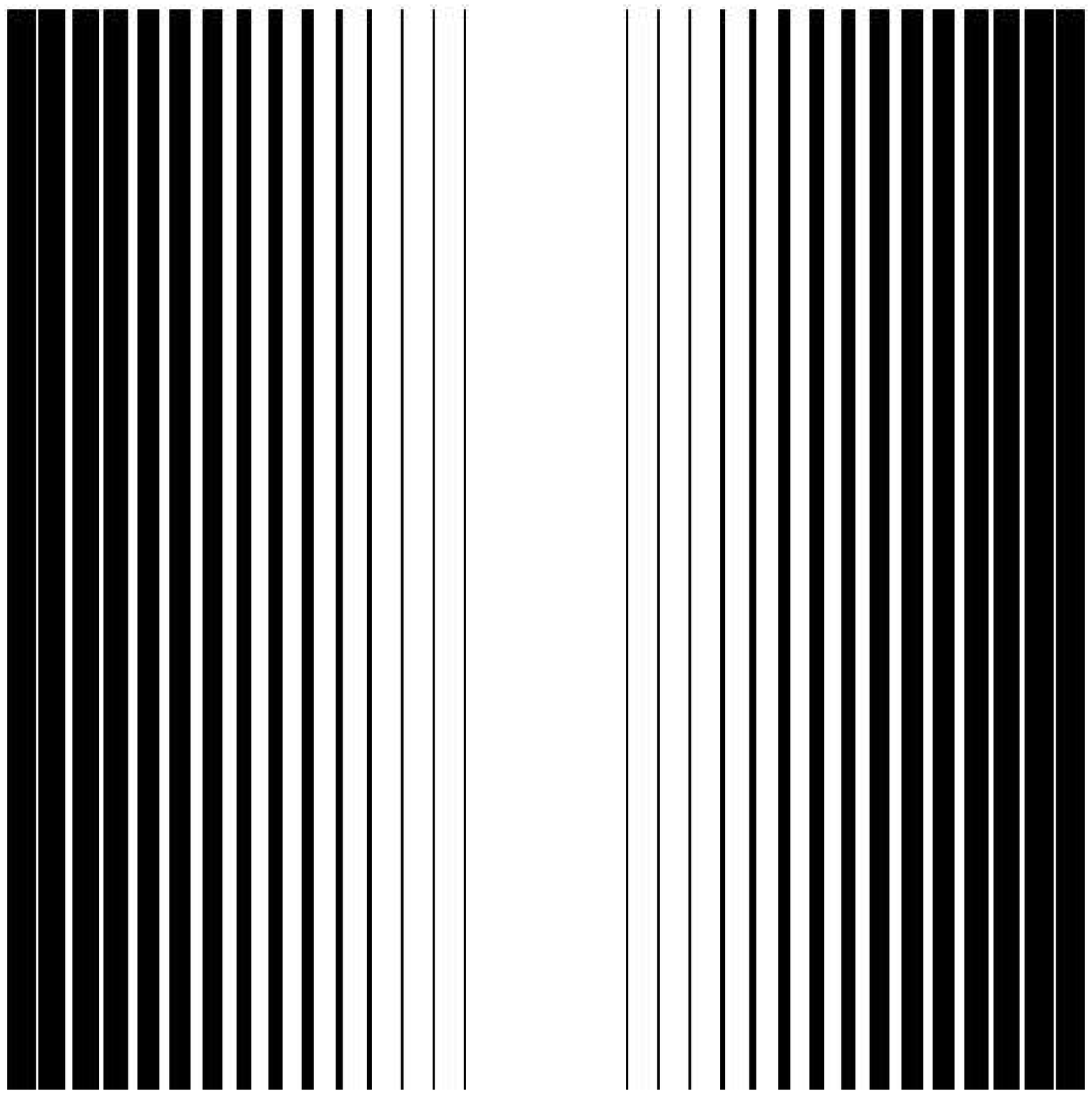}
\includegraphics[width=3.0in]{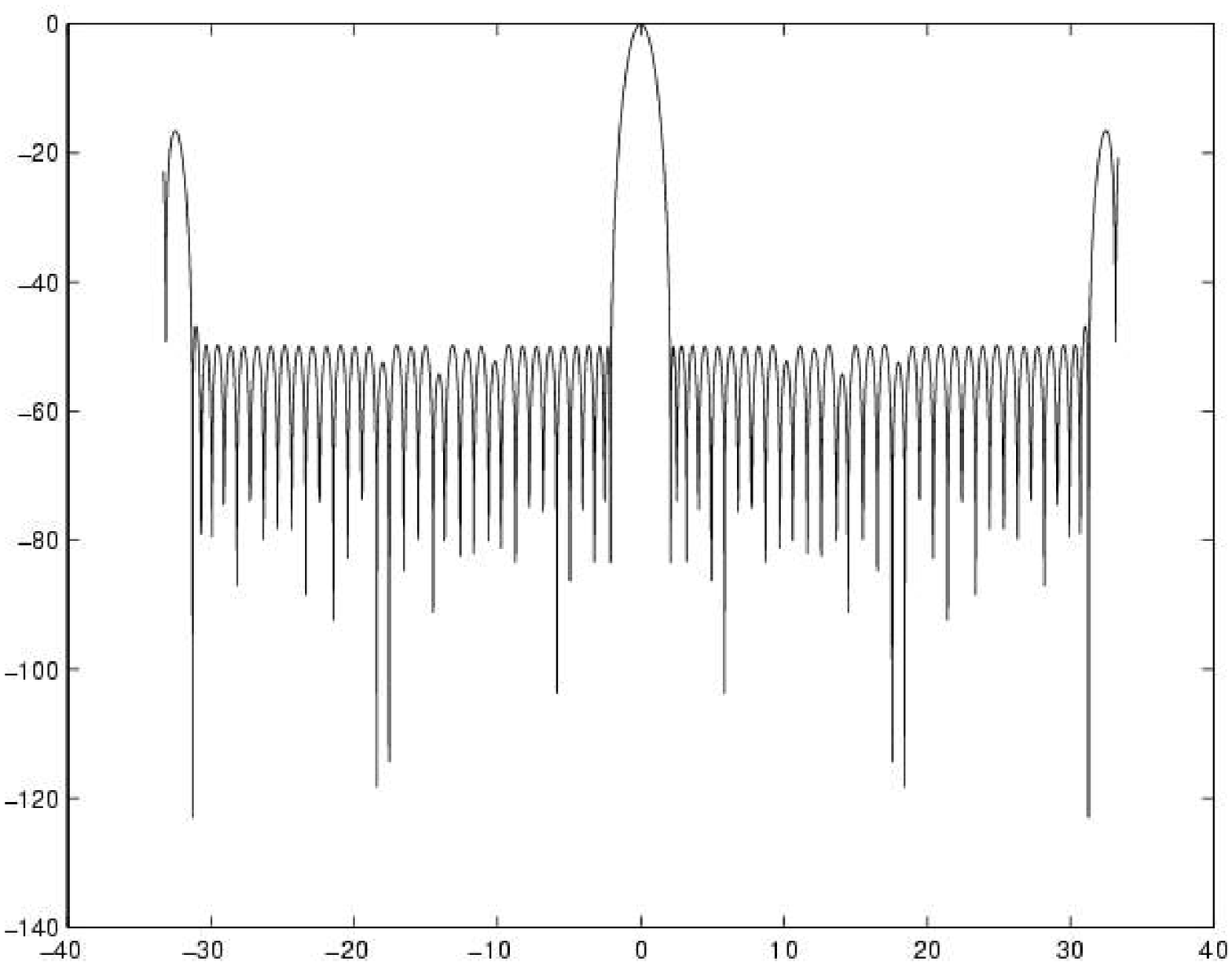}
\\
\includegraphics[width=3.0in]{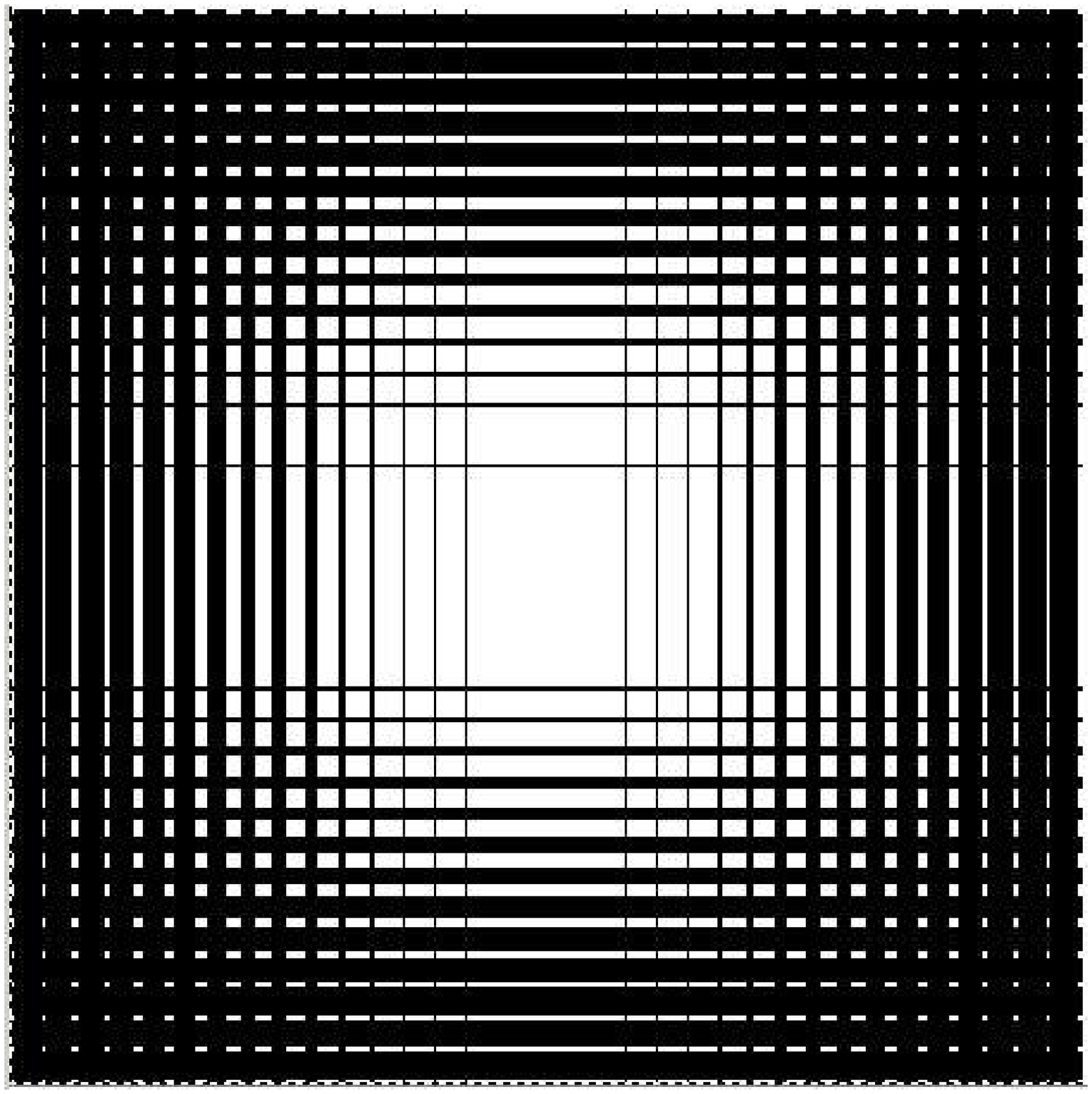}
\includegraphics[width=3.0in]{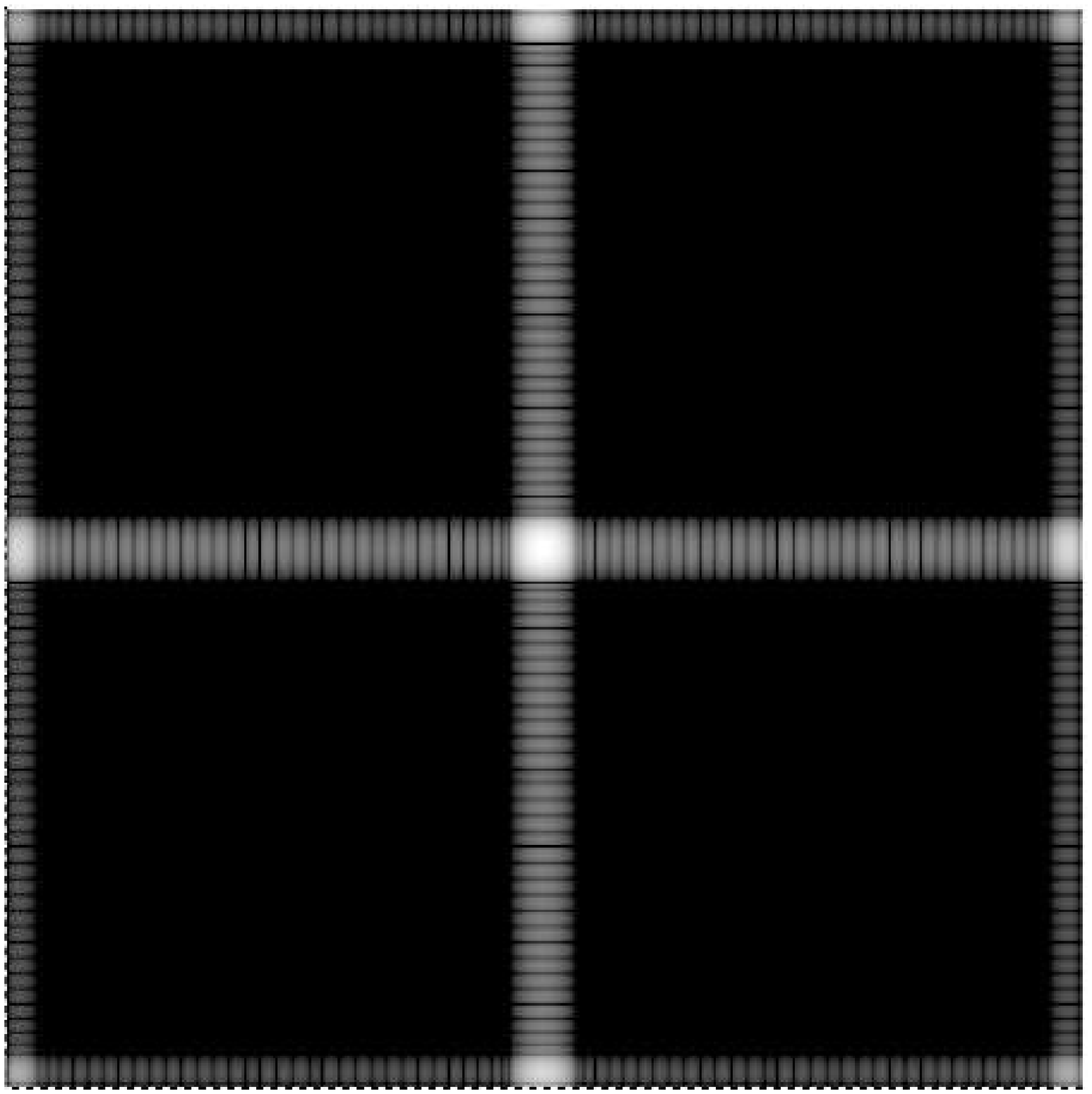}
\end{center}
\caption{
The open area of the rectangular mask is $28.1\%$ 
and the Airy throughput is $15.1\%$. 
{\em Top left.} A barcode mask designed to provide $10^{-5}$ contrast from 
$2 \lambda/D$ to $25 \lambda/D$.
{\em Top right.} The PSF associated with the barcode mask.
{\em Bottom left.} The corresponding rectangular mask.
{\em Bottom right.} The PSF corresponding to the rectangular mask.
The gray-scale represents a logarithmic stretch with black
corresponding to $10^{-10}$ and white corresponding to $1$.
}
\label{fig1}
\end{figure}

The masks considered in this paper consist of rectangular arrays of 
rectangular
opennings.  They can be built as two identical striped masks one placed 
on top of the other but oriented so that the stripes on one mask are orthogonal 
to the stripes on the other (see Figure \ref{fig1}).  
We call the striped masks {\em barcode} masks.
Each barcode mask need only
provide a contrast ratio of $10^{-5}$---it is then guaranteed that the
two-mask overlay will achieve a contrast of $10^{-10}$.  
This is important since it is quite feasible to 
check in a laboratory that a barcode mask achieves $10^{-5}$ 
but is perhaps impossible to check that a mask achieves $10^{-10}$ in any
ground-based laboratory.

Furthermore, we show that one can get an even tighter IWA of 
$1.4\sqrt{2} = 2.0 \lambda/D$ by placing a rectangular pupil-plane mask
in the Lyot plane of a traditional coronagraph and using an image-plane
occulting mask consisting of a ``plus-sign'' shape,
$\{ (\xi,\zeta): \; | \xi | < 0.6, \; | \zeta | < 0.6 \}$.
%
%
We show that such a coronagraph can tolerate pointing errors as large
as about $0.05 \lambda/D$.

Finally, in the interest of producing an on-axis coronagraph, we present a
Lyot-plane rectangular pupil mask which includes a $2\%$ 4-vane spider.
A similar spider can then be placed in the entrance pupil and provide the
opportunity to place a small secondary mirror at the center of the plane.
Having such a spider over the entrance pupil provides the possibility of
building an all-mirror on-axis system and in this way greatly reducing the
negative effects of differential polarization that plague off-axis designs.

\section{Barcode and Rectangular Pupil Masks} \label{sec:barcode}

In this paper we assume that telescope optics follow the Fraunhofer
approximation.  Hence,
given a pupil plane apodization function $0 \le A(x,y) \le 1$, 
the image-plane electric
field is given by the two-dimensional Fourier transform
\begin{equation} \label{1}
    E(\xi,\zeta) = \widehat{A}(\xi,\zeta) 
    := \iint e^{2 \pi i (\xi x + \zeta y)} A(x,y) dy dx.
\end{equation}
If the apodization function takes only the values zero and one, then the
function represents a {\em mask}.  If it depends only on one of the
coordinates (either $x$ or $y$), then we say that it is a
{\em one-dimensional apodization} or a {\em one-dimensional mask}.  
We also refer to one-dimensional masks as {\em barcode} masks.
This family of masks were studied in \citet{KVLS04}.
In this paper, we are interested
in masks that correspond to the tensor product of a pair of one-dimensional
masks:
\begin{equation} \label{2}
    A = A_x \otimes A_y \qquad \Leftrightarrow \qquad A(x,y) = A_x(x) A_y(y) .
\end{equation}
The electric field corresponding to a tensor product is itself a
tensor product:
\begin{equation} \label{3}
    E = \widehat{A} = \widehat{A_x \otimes A_y} 
      = \widehat{A_x} \otimes \widehat{A_y} .
\end{equation}
In other words,
\begin{equation} \label{4}
    E(\xi, \zeta) = \widehat{A_x}(\xi) \widehat{A_y}(\zeta) .
\end{equation}
Tensor products of smooth apodizations were first proposed for terrestrial
planet finding in \cite{ref:Nisenson}.

Since intensity is the square of the magnitude of the electric field, it follows
that for a contrast of $10^{-10}$, we seek apodization functions that provide
the following contrast inequalities:
\begin{equation} \label{5}
    |E(\xi, \zeta)| \le 10^{-5} E(0,0), \qquad (\xi, \zeta) \in {\cal O},
\end{equation}
where ${\cal O}$ denotes the points in the image plane at which high contrast
is to be achieved.  Suppose as before that the apodization function is a 
tensor product.  If the set ${\cal O}$ is a {\em generalized rectangle}
\begin{equation} \label{6}
    {\cal O} = \{ (\xi, \zeta) : \; \xi \in {\cal O}_{\xi}, 
                                 \; \zeta \in {\cal O}_{\zeta} \}
\end{equation}
(i.e., ${\cal O} = {\cal O}_{\xi} \times {\cal O}_{\zeta}$), then the contrast
inequalities can be achieved by giving two one-dimensional apodizations
that each achieve a contrast ratio of only $10^{-5}$:
\begin{eqnarray} \label{7}
    |\widehat{A_x}(\xi  )| & \le & 10^{-2.5} \widehat{A_x}(0), 
	    \qquad \xi   \in {\cal O}_{\xi},
    \\
    |\widehat{A_y}(\zeta)| & \le & 10^{-2.5} \widehat{A_y}(0),
	    \qquad \zeta \in {\cal O}_{\zeta}.
\end{eqnarray}

Figure \ref{fig1} shows a rectangular mask corresponding to
${\cal O}_{\xi} = {\cal O}_{\zeta} = \{ \xi : 2 \le | \xi | \le 25 \}$.
The mask has a $28.1\%$ open area.  Its {\em Airy throughput},
defined as 
\begin{equation} \label{8}
    \int_{-\xi_0}^{\xi_0} \int_{-\xi_0}^{\xi_0}
    | E(\xi, \zeta) |^2 d\zeta d\xi,
\end{equation}
where $\xi_0$ denotes the inner working angle for the barcode mask,
is $15.1\%$.

\begin{figure}
\begin{center} 
\includegraphics[width=3.0in]{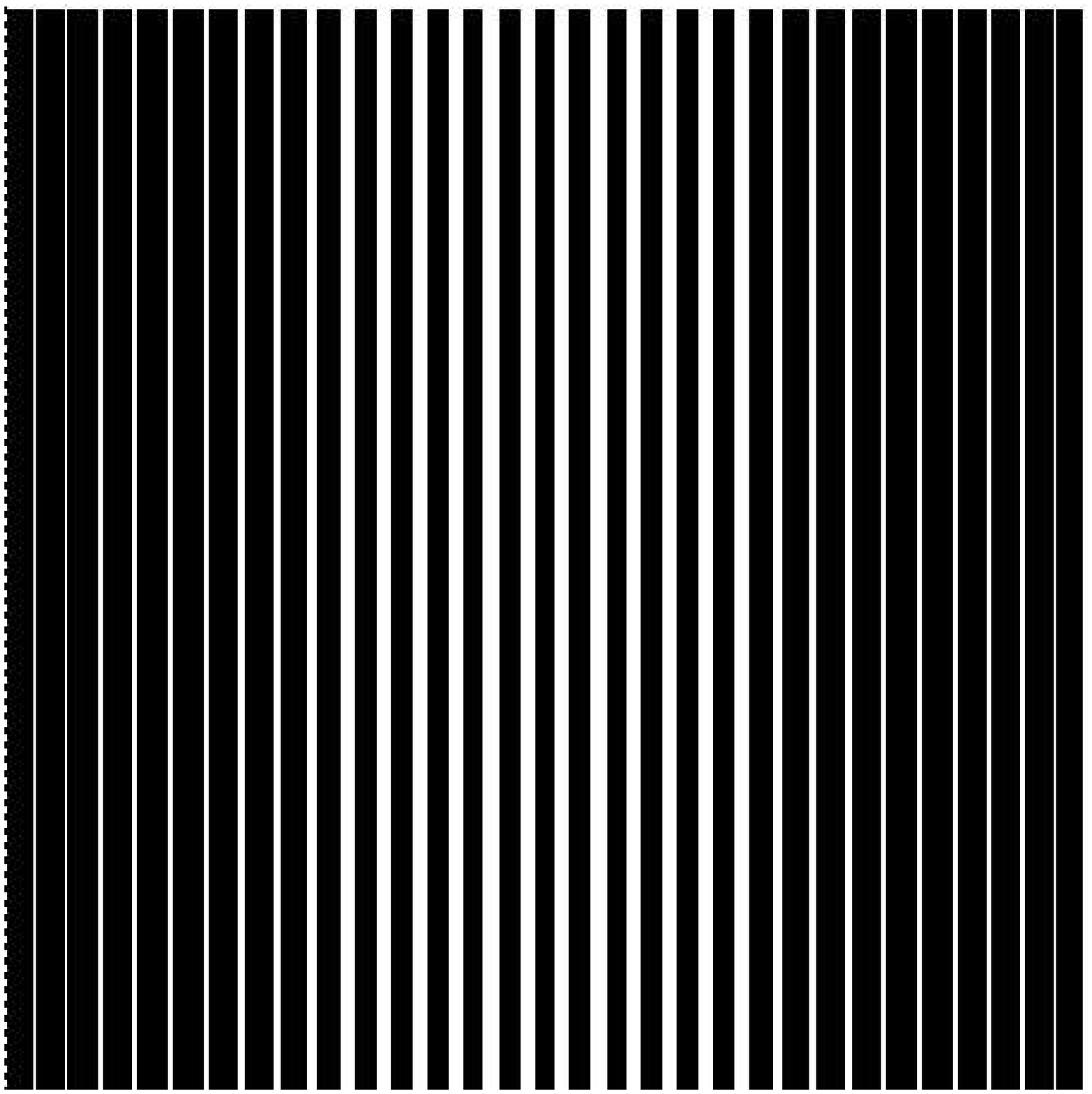}
\includegraphics[width=3.0in]{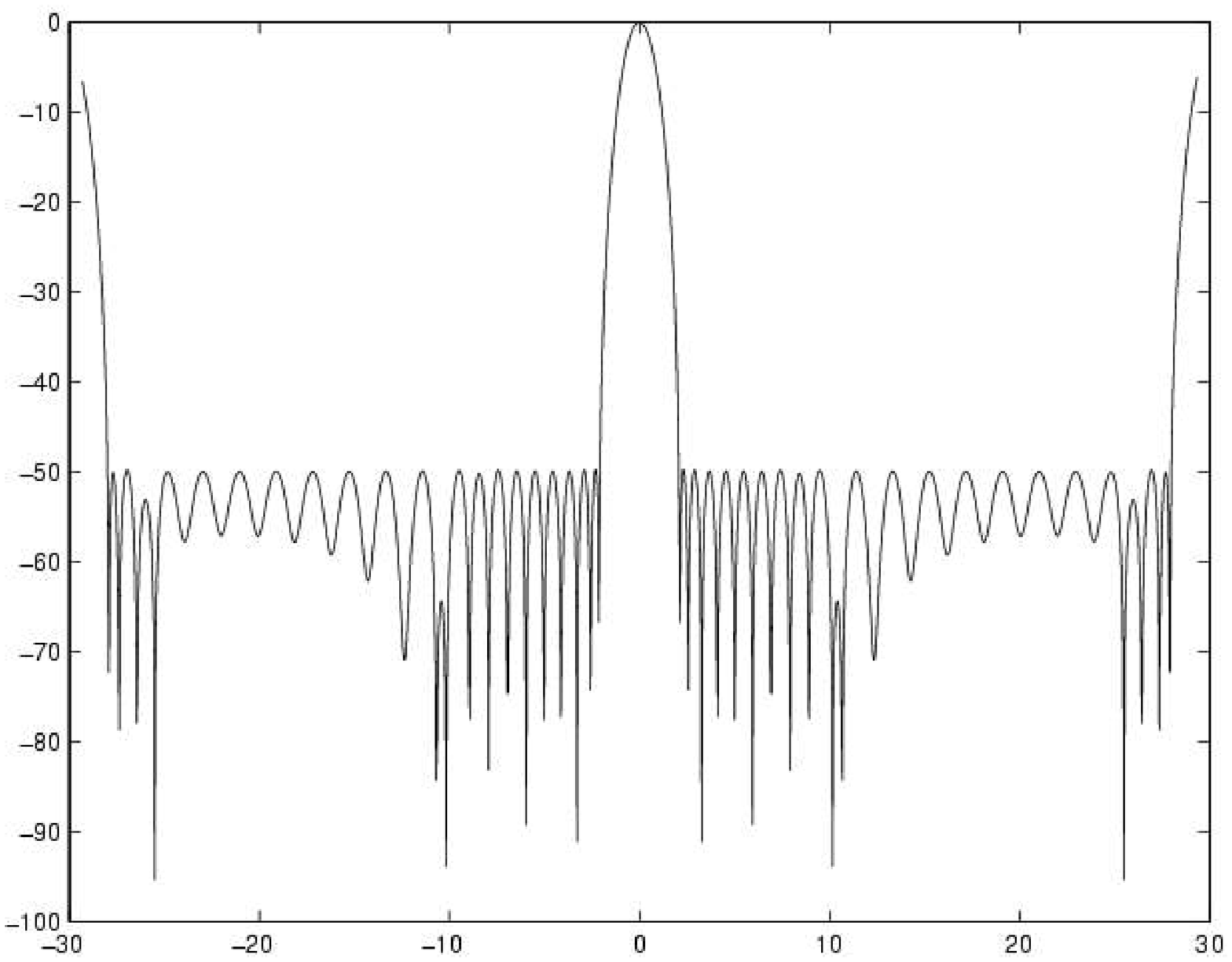}
\\
\includegraphics[width=3.0in]{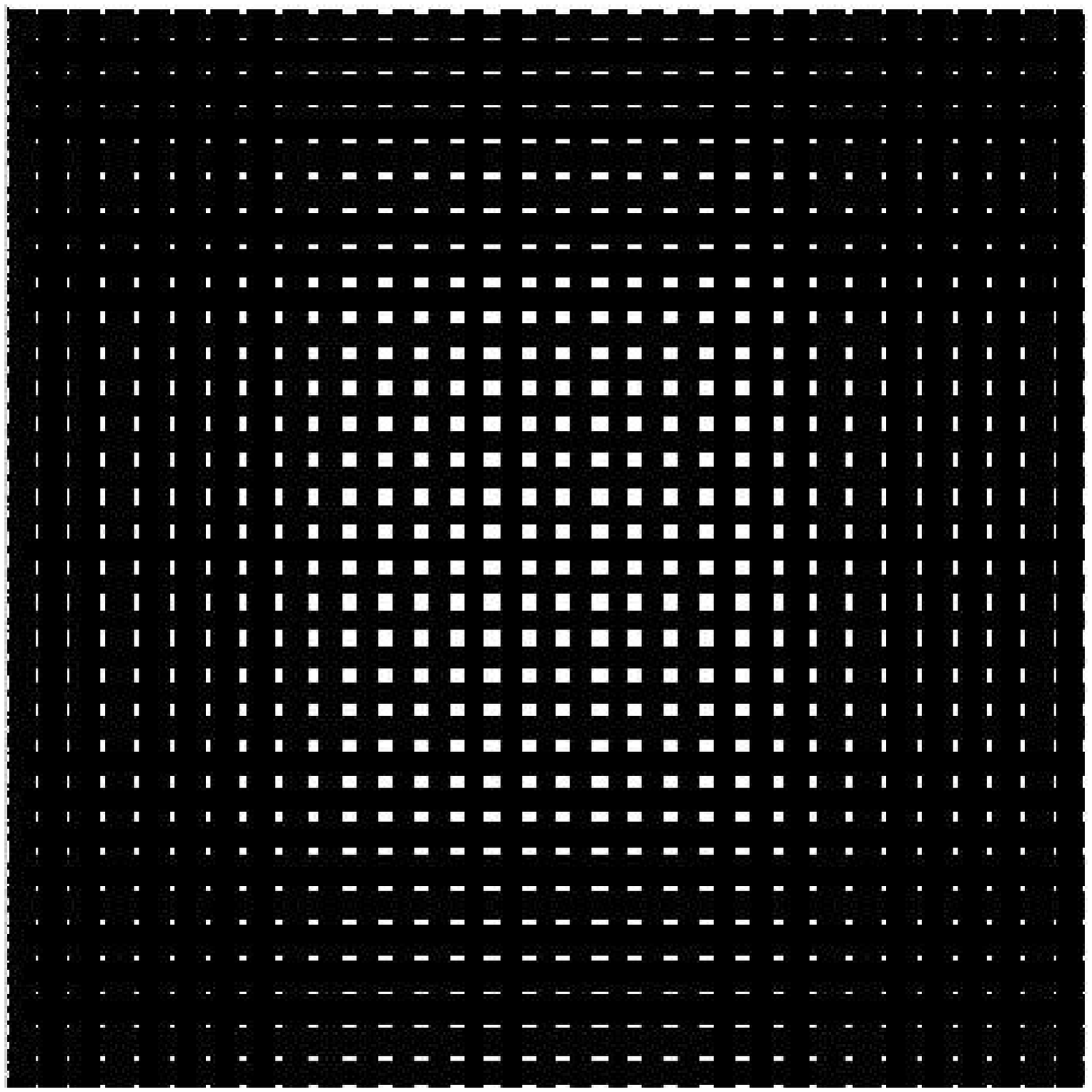}
\includegraphics[width=3.0in]{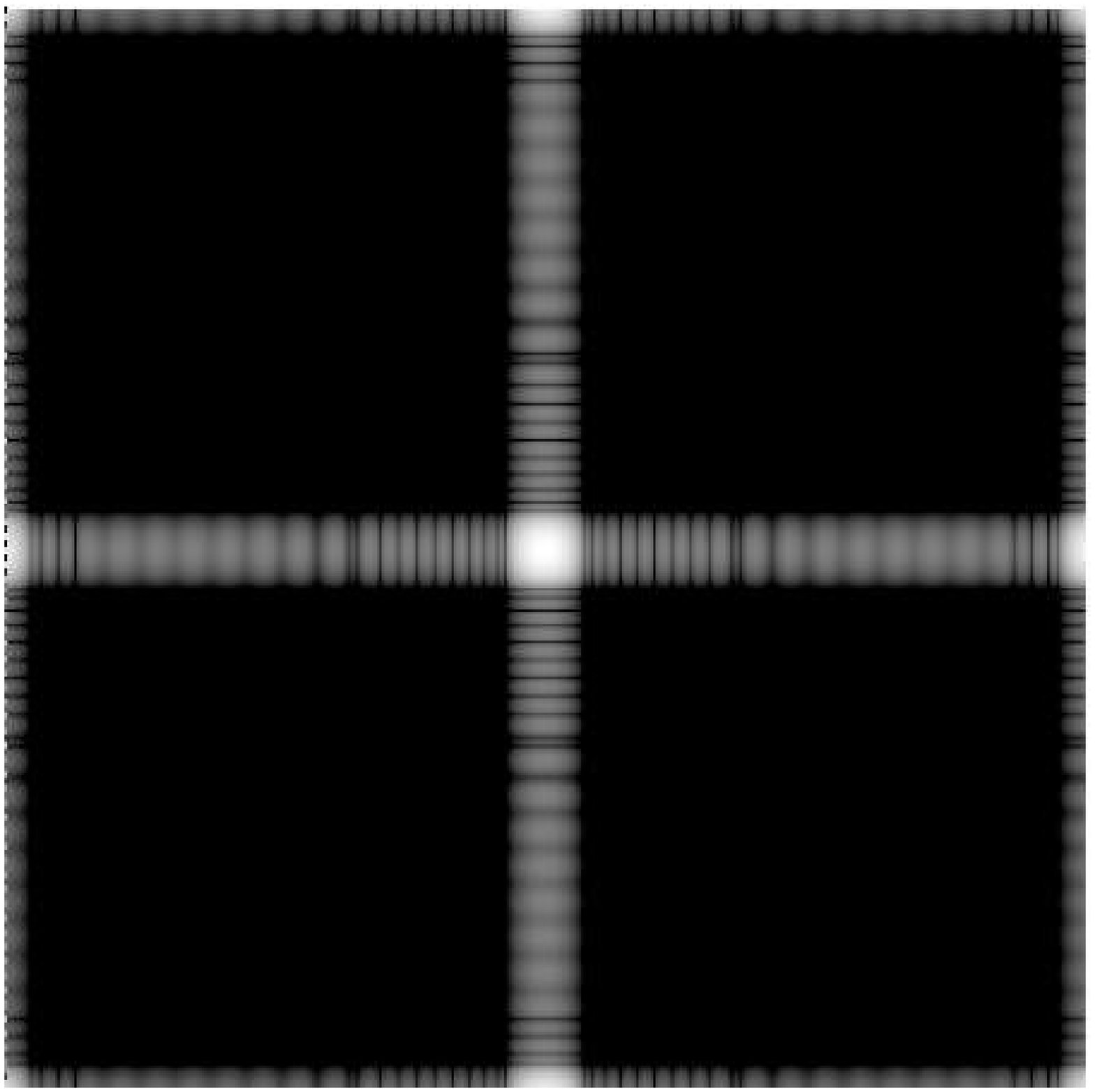}
\end{center}
\caption{
This design includes a $2\%$ central stripe to hang/hide a secondary
mirror.  
The open area of the rectangular mask is $4.6\%$ 
and the Airy throughput is $0.41\%$. 
{\em Top left.} A barcode mask designed to provide $10^{-5}$ contrast from 
$2 \lambda/D$ to $22 \lambda/D$.
{\em Top right.} The PSF associated with the barcode mask.
{\em Bottom left.} The corresponding rectangular mask.
{\em Bottom right.} The PSF corresponding to the rectangular mask.
The gray-scale represents a logarithmic stretch with black
corresponding to $10^{-10}$ and white corresponding to $1$.
}
\label{fig2}
\end{figure}

Figure \ref{fig2} shows another rectangular mask.  For this mask,
a $2\%$ central obstruction was imposed on the design.  With such
a central obstruction (and implied spiders) one can hang and hide
a small secondary mirror and therefore build a telescope with an
on-axis optical path.  In this design, we have used
${\cal O}_{\xi} = {\cal O}_{\zeta} = \{ \xi : 2 \le | \xi | \le 22 \}$.
The mask has a $4.6\%$ open area.  Its {\em Airy throughput},
is $0.41\%$, which is somewhat disappointing.  In the Section \ref{sec4},
we present an on-axis design that achieves a much higher throughput.

We end this section by point out that PSFs associated with these masks 
depend on wavelength only in that the inner working angle is measured
in units of $\lambda/D$, where $\lambda$ is wavelength and $D$ is aperture.
Hence, measured in radians, at longer wavelengths the inner working angle
is correspondingly larger.

\section{A Lyot Coronagraph}

Consider an imaging system consisting of an entrance pupil,
a first image plane, a reimaged (Lyot) pupil plane, and a final
image plane (containing an imaging device).  We assume that
each of the first three planes can be apodized/masked.  
Following \cite{ref:kuchner}, we let $A$ denote
the mask function for the entrance pupil, $\hat{M}$ 
the mask function for the first image plane, and $L$ 
the mask function for the Lyot pupil.  

The electric field at point $(\xi,\zeta)$ 
in the final image plane corresponding to an off-axis point source 
is a composition of mask multiplication and Fourier transform:
\begin{equation} \label{10}
    {\cal F} \left( L \cdot 
        {\cal F} \left( \widehat{M} \cdot
            {\cal F} \left( A \cdot F_{-\xi_0,-\zeta_0} \right)
	\right)
    \right)(\xi, \zeta)
    =
    \widehat{L}*
    \left( \widehat{M} \cdot (\widehat{A} * \delta_{\xi_0,\zeta_0}) \right)
    (\xi, \zeta)
    ,
\end{equation}
where 
$ F_{-\xi_0,-\zeta_0} $ denotes the electric field at the entrance pupil
corresponding to a point source from direction $(\xi_0,\zeta_0)$,
$\delta_{\xi_0,\zeta_0}$ denotes a unit mass delta function at
$(\xi_0,\zeta_0)$, hats denote Fourier transforms and stars denote
convolutions.

It is easy to see that if each of $L$, $M$, and $A$ are tensor products
($L = L_x \otimes L_y$, $M = M_x \otimes M_y$, $A = A_x \otimes A_y$),
then the electric field again factors into a product of electric fields:
\begin{equation} \label{12}
    \widehat{L}*
    \left( \widehat{M} \cdot (\widehat{A} * \delta_{\xi_0,\zeta_0}) \right)
    (\xi, \zeta)
    =
    \widehat{L_x}*
    \left( \widehat{M_x} \cdot (\widehat{A_x} * \delta_{\xi_0}) \right)
    (\xi)
    \quad
    \widehat{L_y}*
    \left( \widehat{M_y} \cdot (\widehat{A_y} * \delta_{\zeta_0}) \right)
    (\zeta)
\end{equation} 
We use this factorization to greatly simplify the coronagraph
optimization problems described in the next section.

\section{Lyot Coronagraph Optimization Problem and Numerical Results}
\label{sec4}

As in our previous papers (\cite{KVSL02,VSK02,VSK03}), 
we use numerical optimization to maximize
some measure $\theta$ of throughput subject to imposed contrast constraints.
For the designs presented in this section, we solve the following
one-dimensional barcode optimization problem:

\begin{equation} \label{9}
\begin{array}{rrlll}
    \text{maximize } & \theta \\[0.2in]
    \text{subject to } &
    \theta 
    & \le 
    \widehat{L}*
    \left( \widehat{M} \cdot (\widehat{A} * \delta_{\xi,\zeta}) \right)
    (\xi,\zeta), 
    & \quad (\xi,\zeta) \in {\cal O},
    \\[0.2in]
    &
    \left|
      \widehat{L}*
      \left( \widehat{M} \cdot (\widehat{A} * \delta_{0}) \right) (\xi,\zeta)
    \right|
    & \le 
      \displaystyle
      \frac{
        \widehat{L}*
        \left( \widehat{M} \cdot (\widehat{A} * \delta_{\xi,\zeta}) \right) (\xi,\zeta)
      }{
        10^{2.5}
      }, 
      & \quad (\xi,\zeta) \in {\cal O}.
\end{array}
\end{equation}
The first set of constraints say that $\theta$ is a lower bound on the
electric field of a unit off-axis source (i.e, a planet as bright as a star)
as the off-axis source varies over the high-contrast region.
The second set of constraints say that the magnitude of the
electric field at $\xi \in {\cal O}$ due to an
on-axis star is smaller than the electric field at the same point in the image
plane due to a much fainter planet whose image is centered at this point $\xi$.
The functions $L$, $M$, and $A$ are functions of a single real variable
and represent either the $x$ or the $y$ component of a tensor product that one
forms to make the final two dimensional image.
By specifying prescribed functions for $A$ and $M$ (actually, $\widehat{M}$)
we can optimize only the function $L$.  In this case, the problem is an
infinite dimensional linear programming problem.  Discretizing the image
and pupil planes reduces the problem to a finite dimensional linear
programming problem that can be solved numerically.

\begin{figure}
\begin{center} 
\includegraphics[width=2.0in]{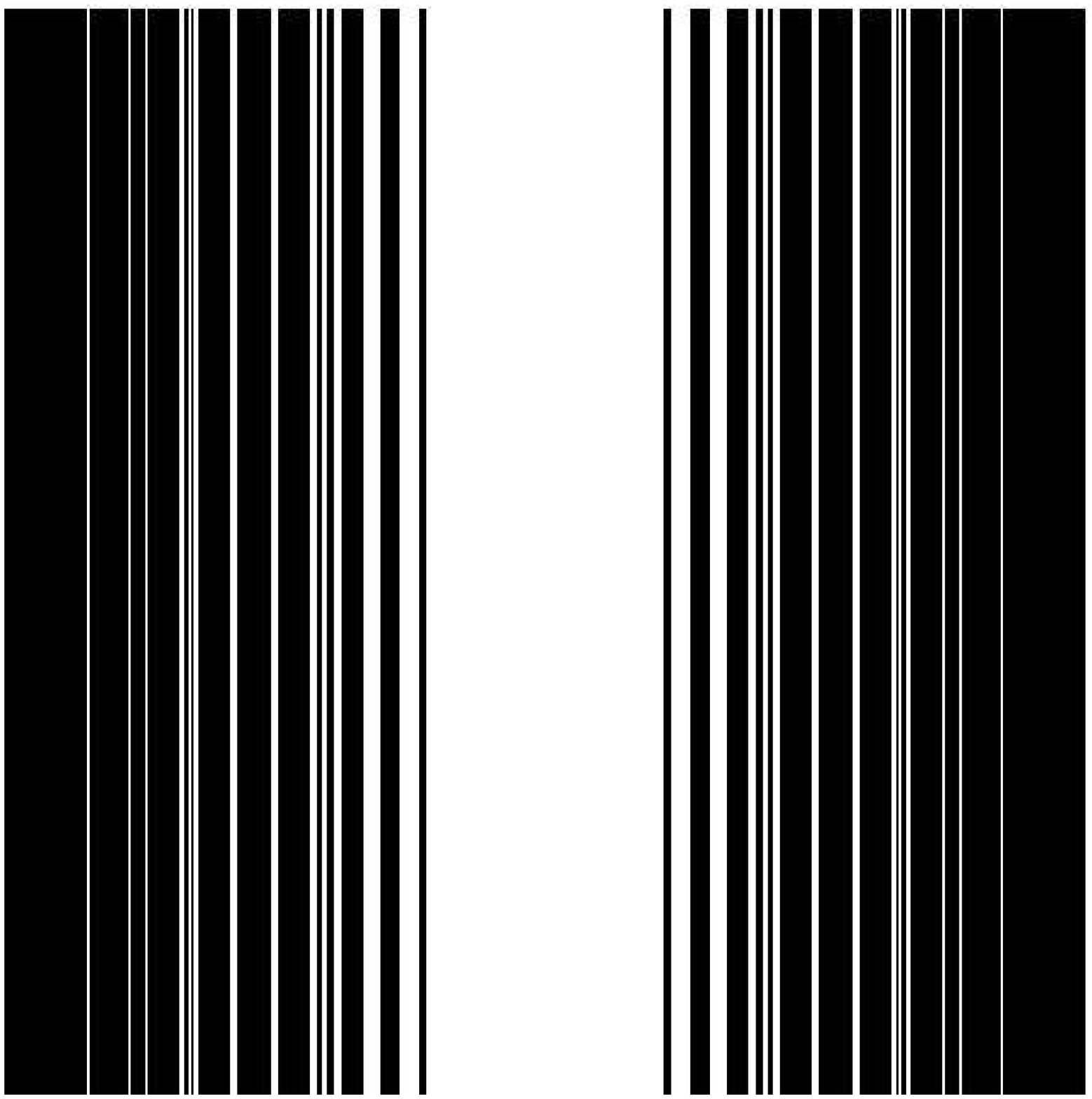}
\includegraphics[width=2.0in]{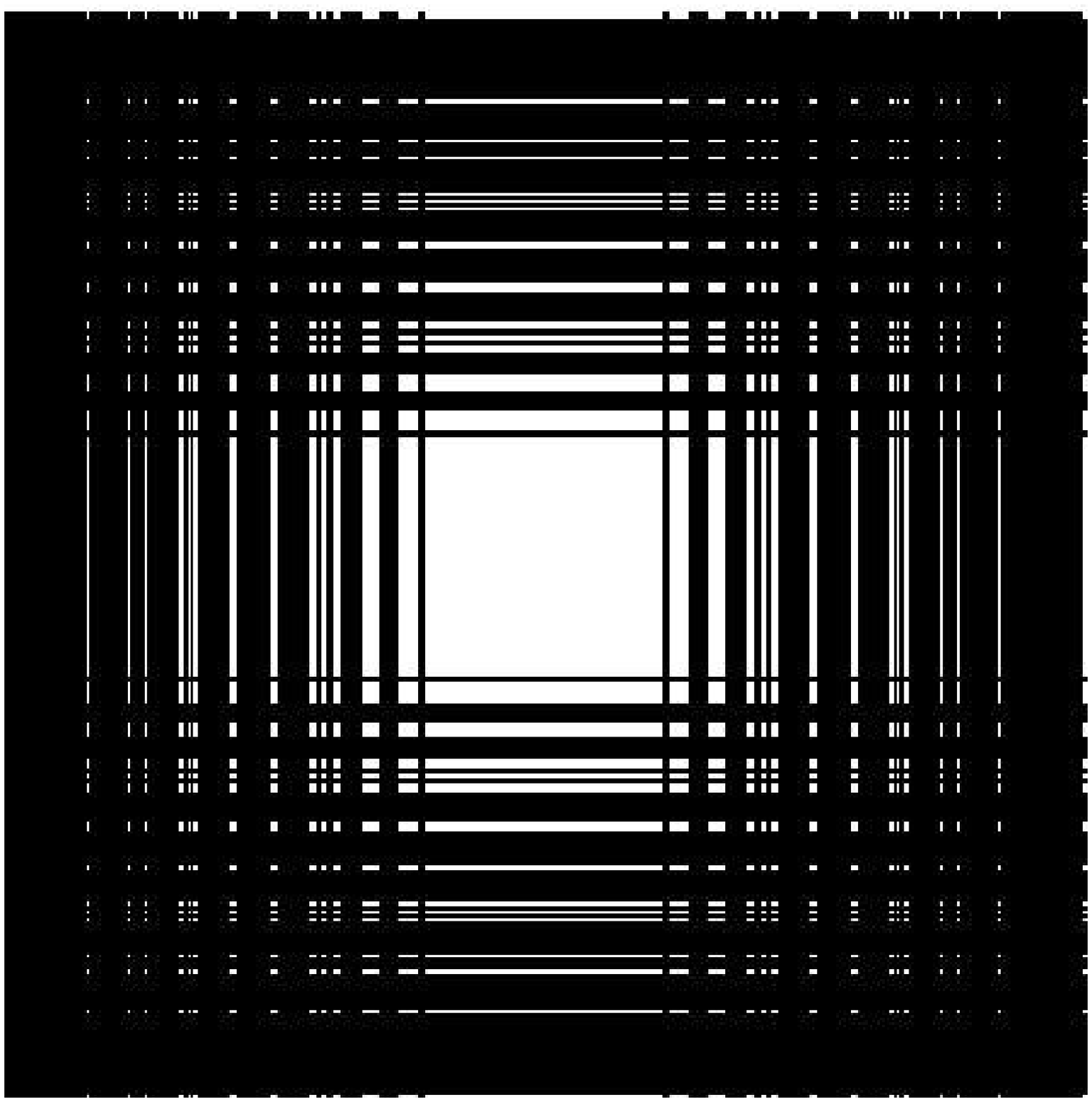}
\includegraphics[width=2.0in]{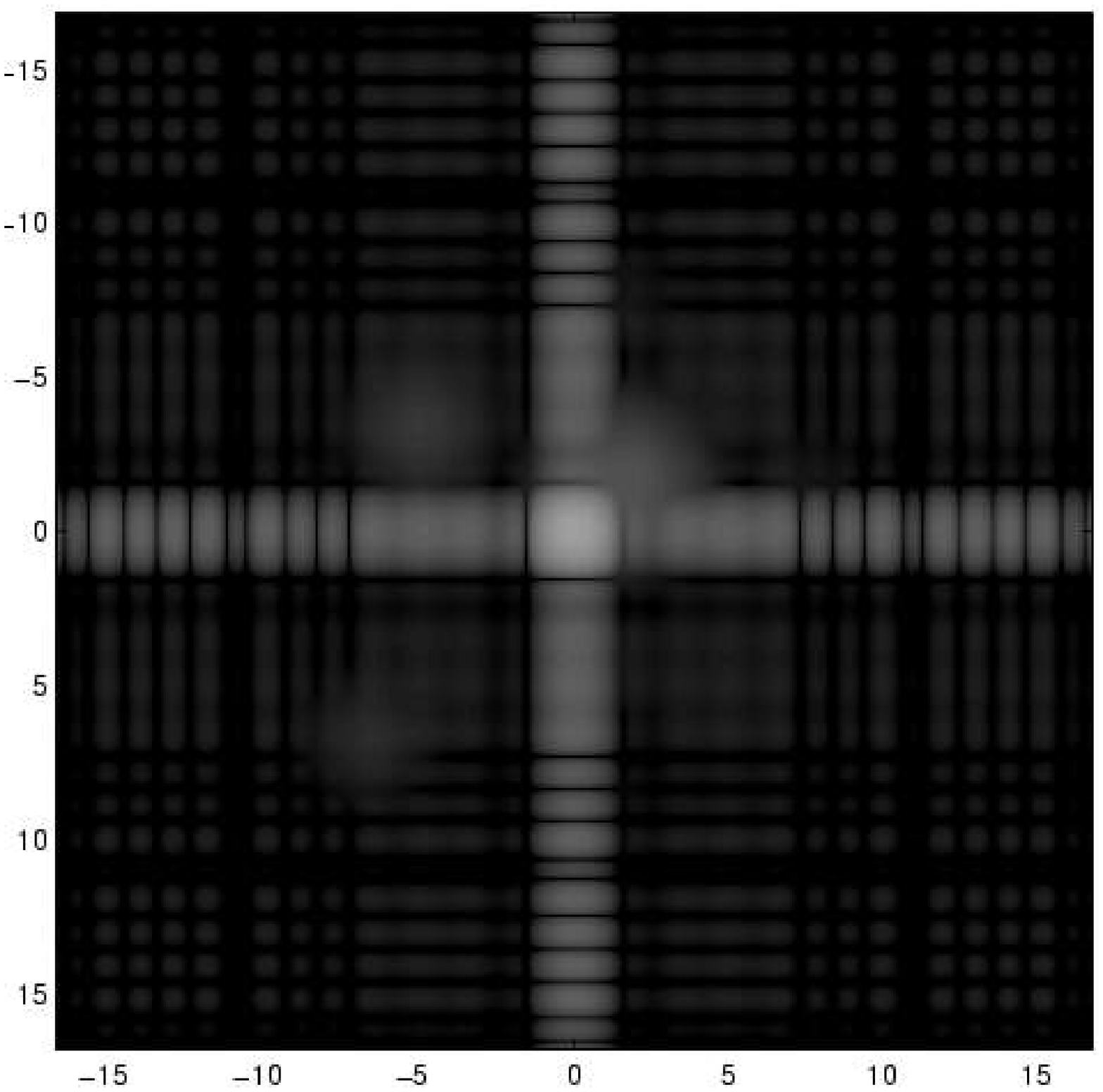}
\\
\includegraphics[width=2.0in]{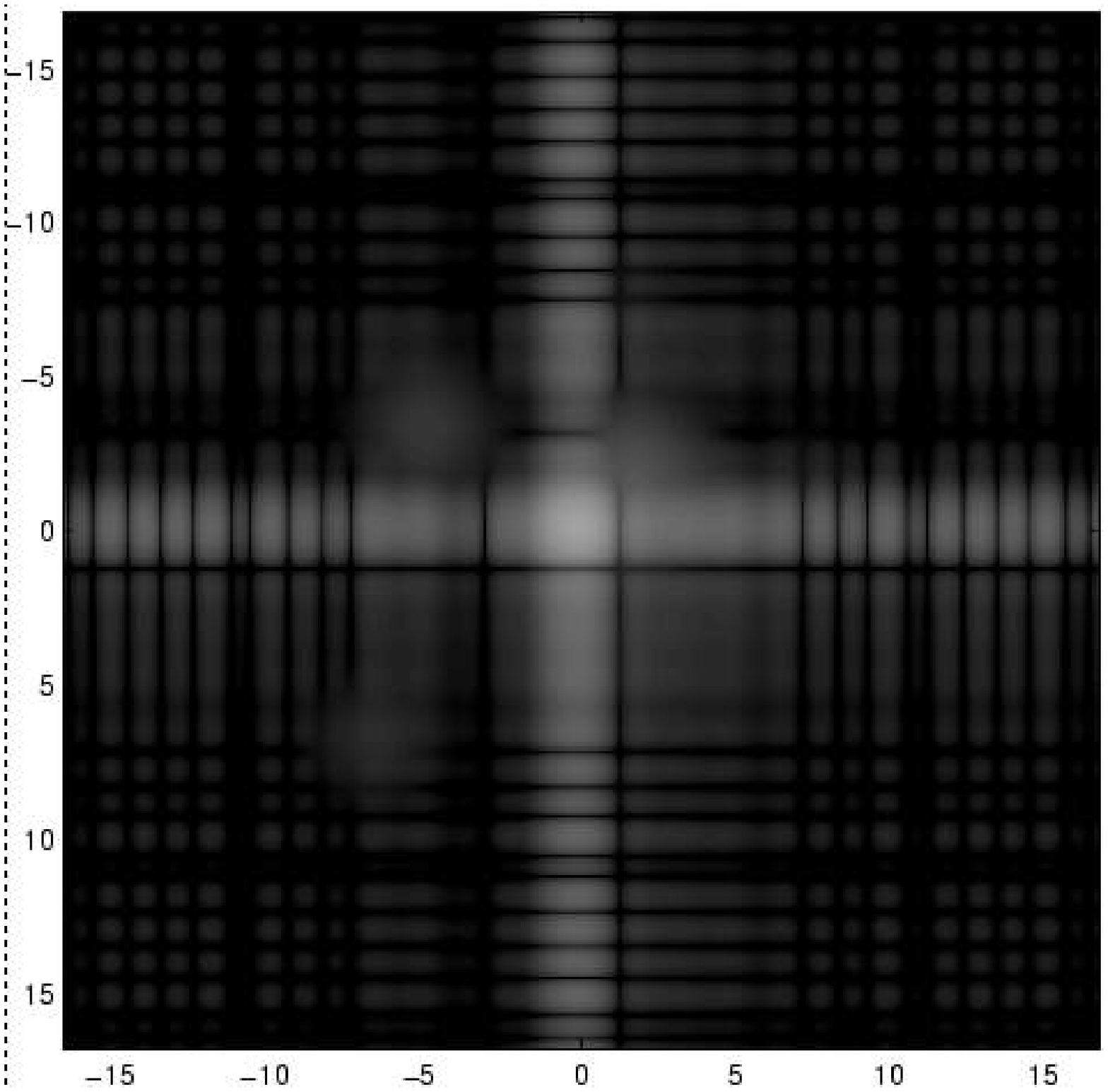}
\includegraphics[width=2.0in]{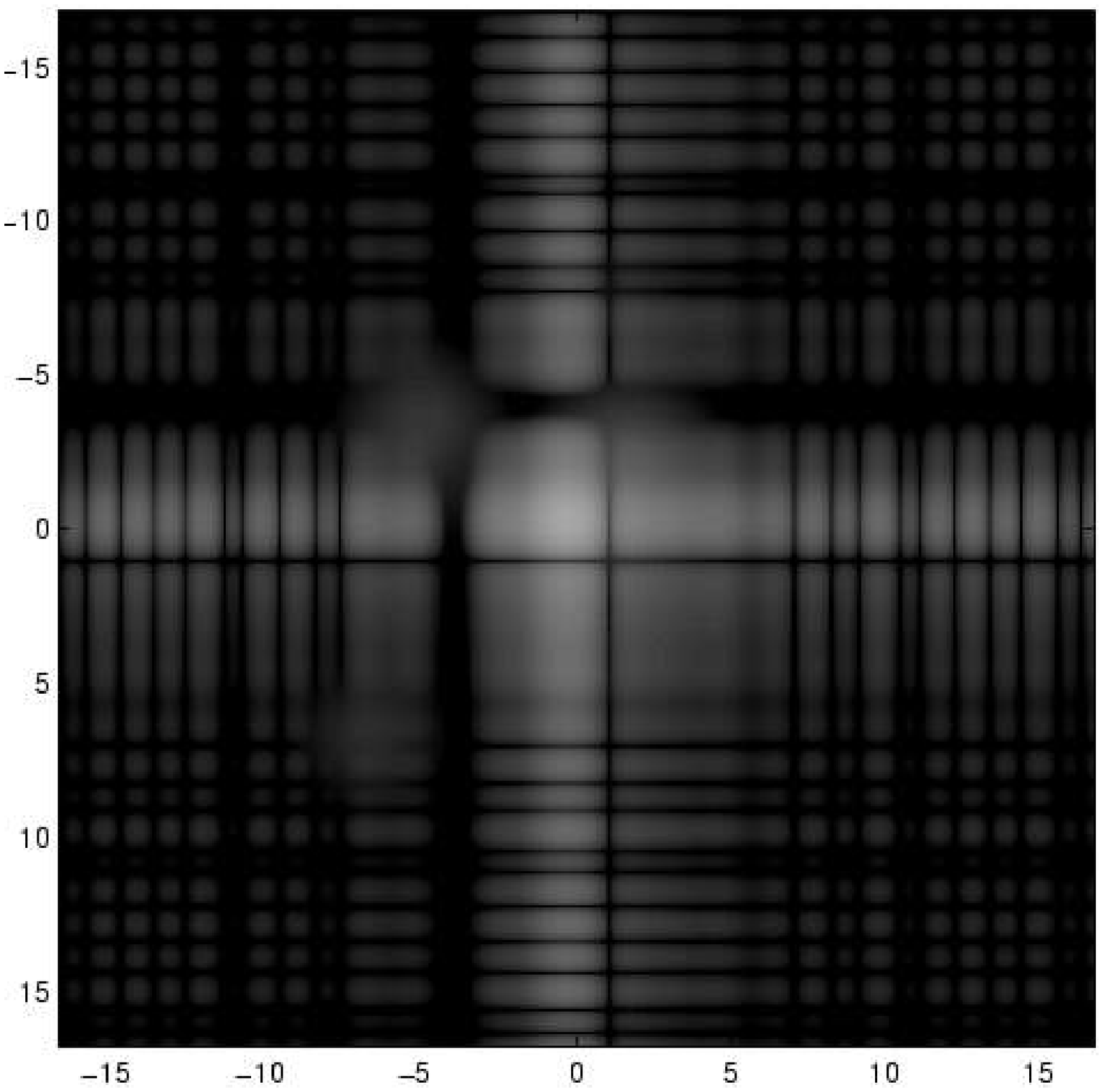}
\includegraphics[width=2.0in]{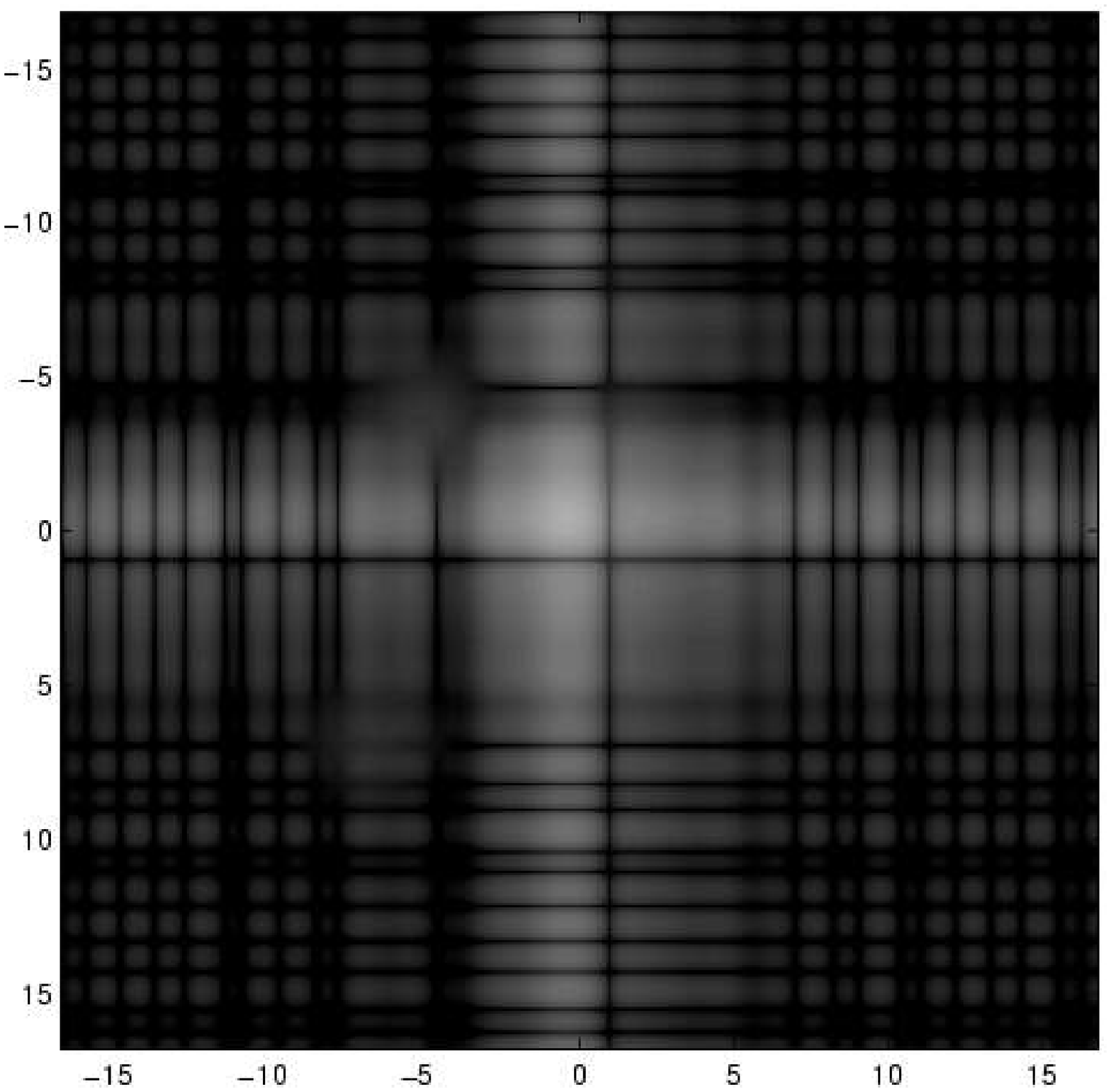}
\end{center}
\caption{
A simulated 3-planet star system as it would appear in a Lyot-style
coronagraph in which the entrance pupil is an open square,
the first image
plane has a ``plus'' shaped occulter of half-width $0.6 \lambda/D$, and a
rectangular mask in the Lyot plane.
{\em Top left.} A barcode mask designed to provide $10^{-5}$ contrast from 
$1.4 \lambda/D$ to $21 \lambda/D$ when used in conjunction with the
image-plane occulter.
{\em Top middle.} The corresponding rectangular mask.
The Airy throughput for this mask is $6.2\%$.  
{\em Top right.} The simulated image.  
The planet in the upper right quadrant is
located at $(1.7,1.7)$ and is $10^{-8}$ times as bright as the star.
The planet in the upper left quadrant is
located at $(-5.1,3.4)$ and is $10^{-9}$ times as bright as the star.
The planet in the lower left quadrant is
located at $(-6.8,-6.8)$ and is $3 \times 10^{-10}$ times as bright as the star.
The gray-scale image is logarithmically stretched to highlight the faint
planets. 
{\em Bottom.} The effect of pointing error.  From left to right, the pointing
error is $0.048$, $0.096$, and $0.144$.  The direction of the error is along
the diagonal into the first quadrant.
}
\label{fig3}
\end{figure}

We now give two specific examples of Lyot coronagraphs.
For the first example, we specify that
\begin{eqnarray} 
    \label{13}
    A(x) = \left\{ \begin{array}{ll} 
     		1 & |x| \le \half \\
		0 & \text{otherwise} 
           \end{array} \right. 
    \\[0.1in]
    \label{14}
    \widehat{M}(\xi) = \left\{ \begin{array}{ll} 
			1 & 0.6 \le | \xi | \le 84 \\
			0 & \text{otherwise} 
                       \end{array} \right. 
    \\[0.1in]
    \label{18}
    |L(x)| \quad \left\{ \begin{array}{ll} 
     		\le 1 & |x| \le \half \\
		 =  0 & \text{otherwise} 
           \end{array} \right. 
    \\[0.1in]
    \label{15}
    {\mathcal O} = \{ \xi : 1.4 \le | \xi | \le 21 \}
\end{eqnarray}
The optimal barcode mask is shown in Figure \ref{fig3} together with
the corresponding rectangular mask.

\begin{figure}
\begin{center} 
\includegraphics[width=2.0in]{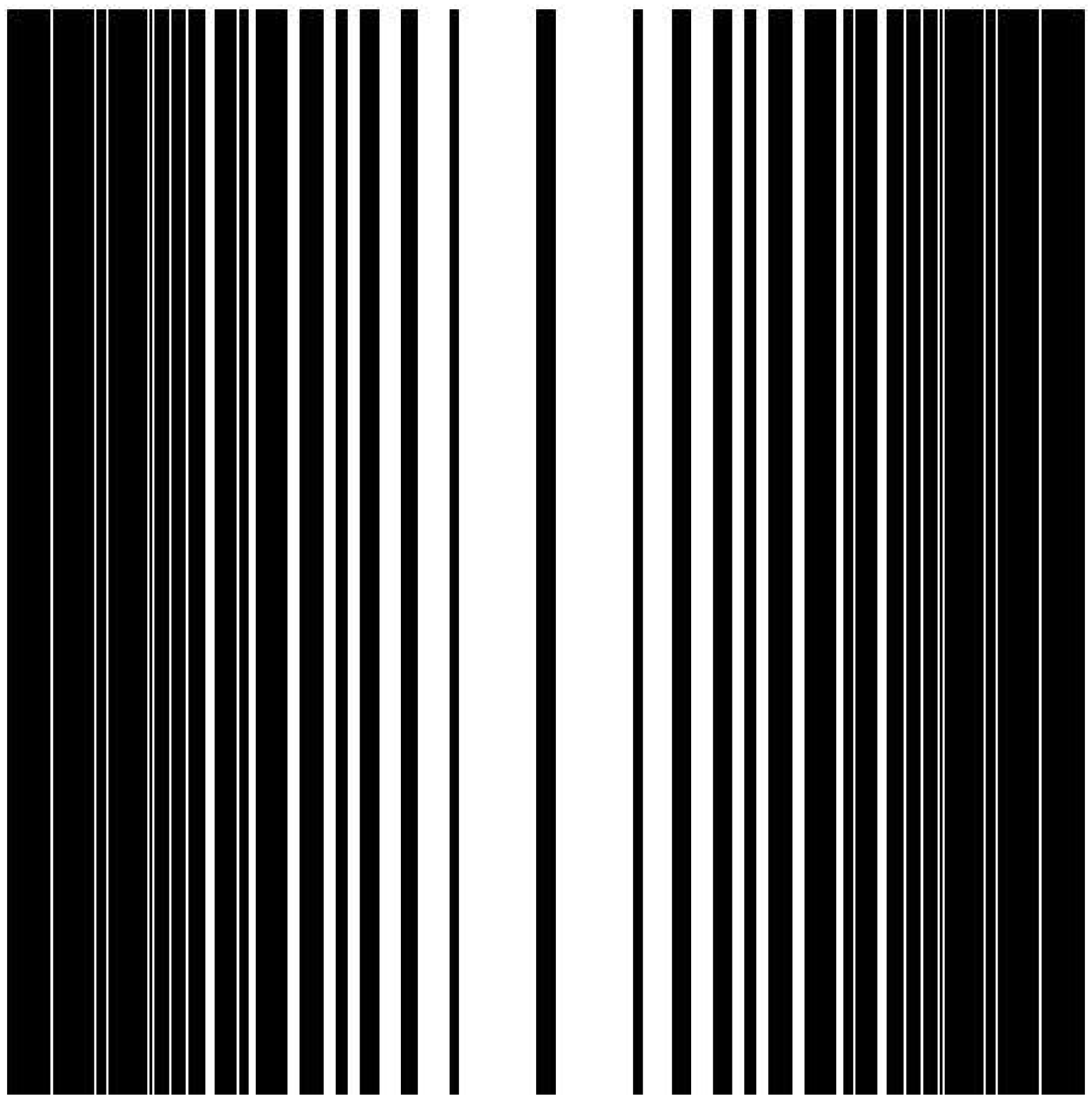}
\includegraphics[width=2.0in]{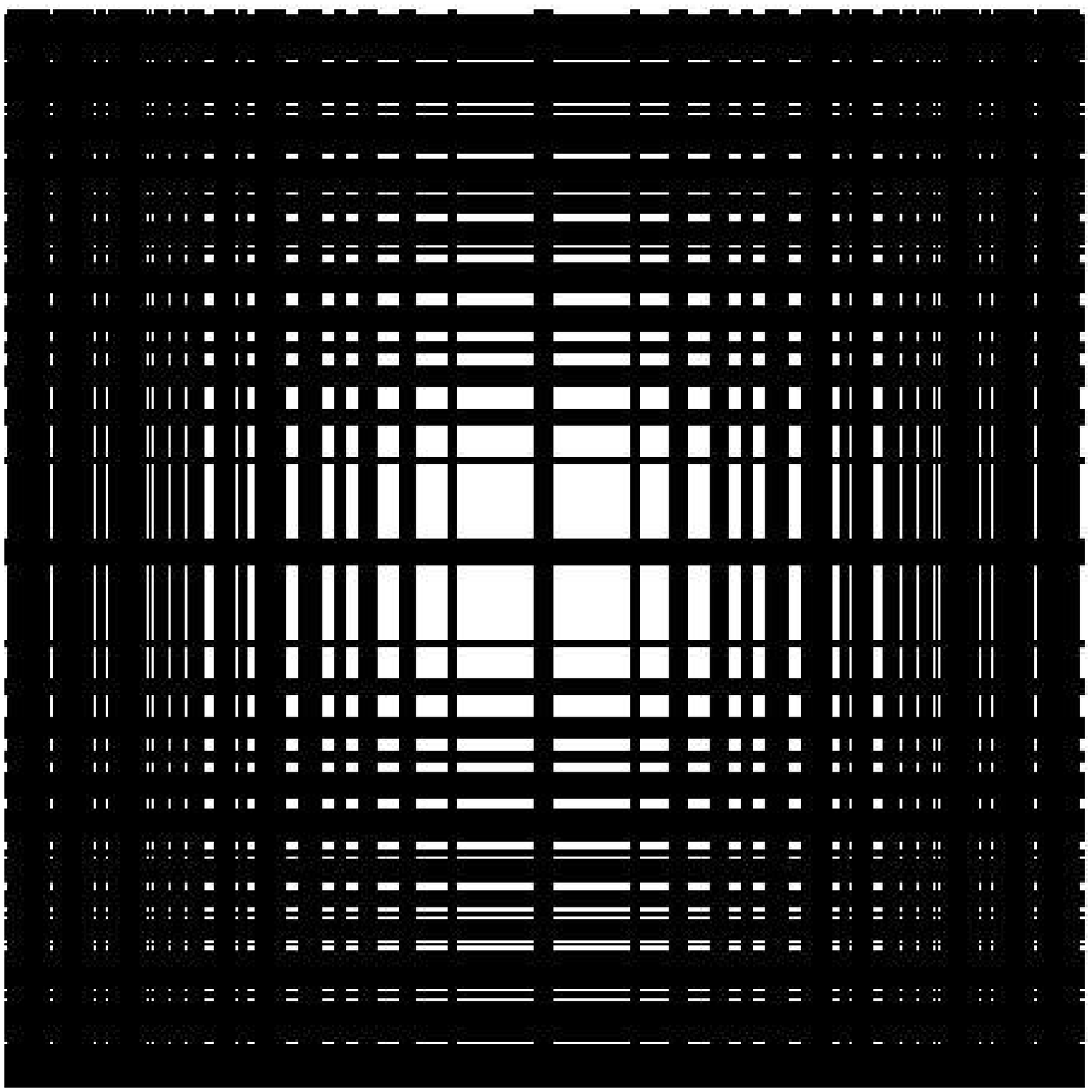}
\includegraphics[width=2.0in]{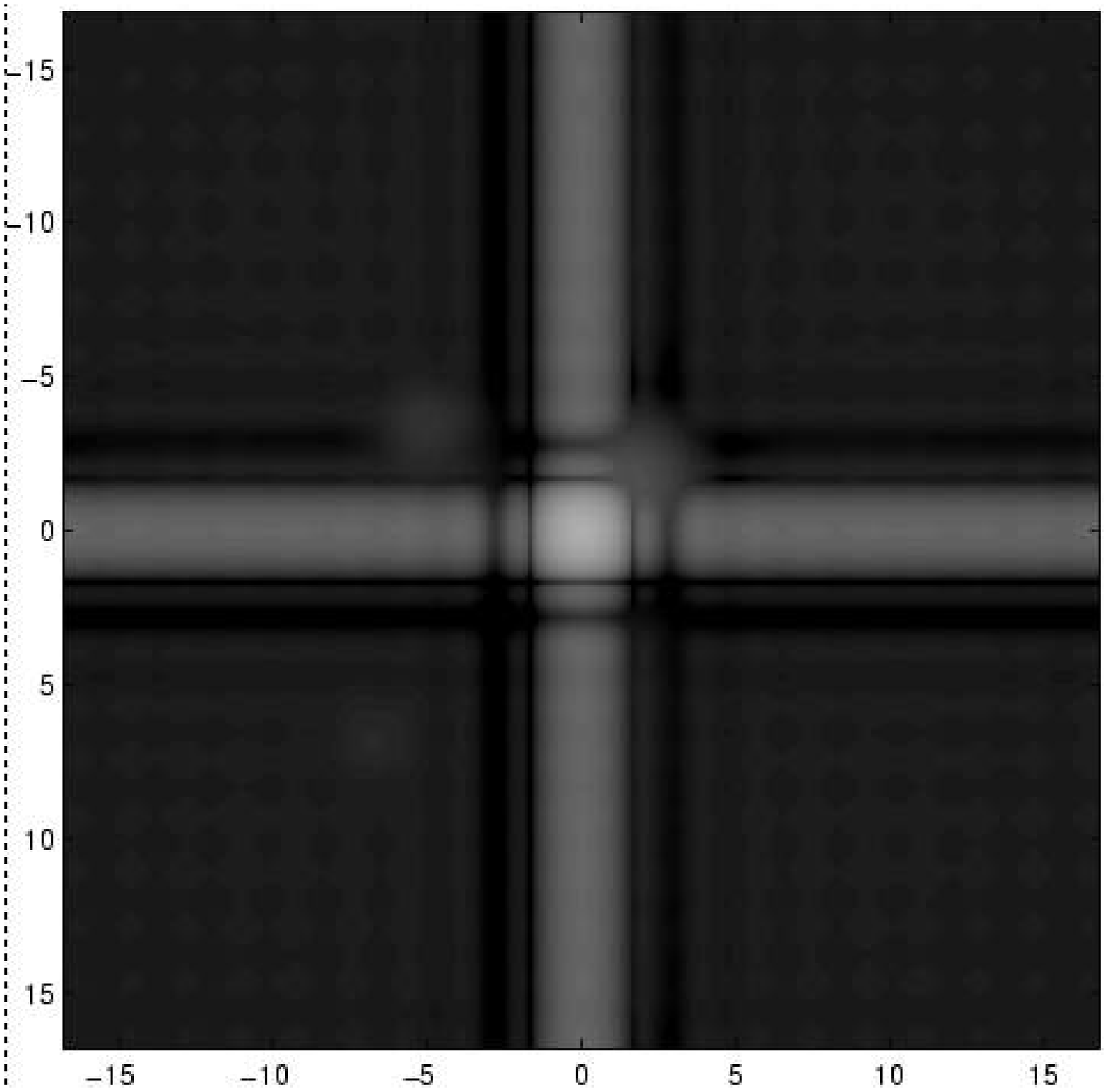}
\\
\includegraphics[width=2.0in]{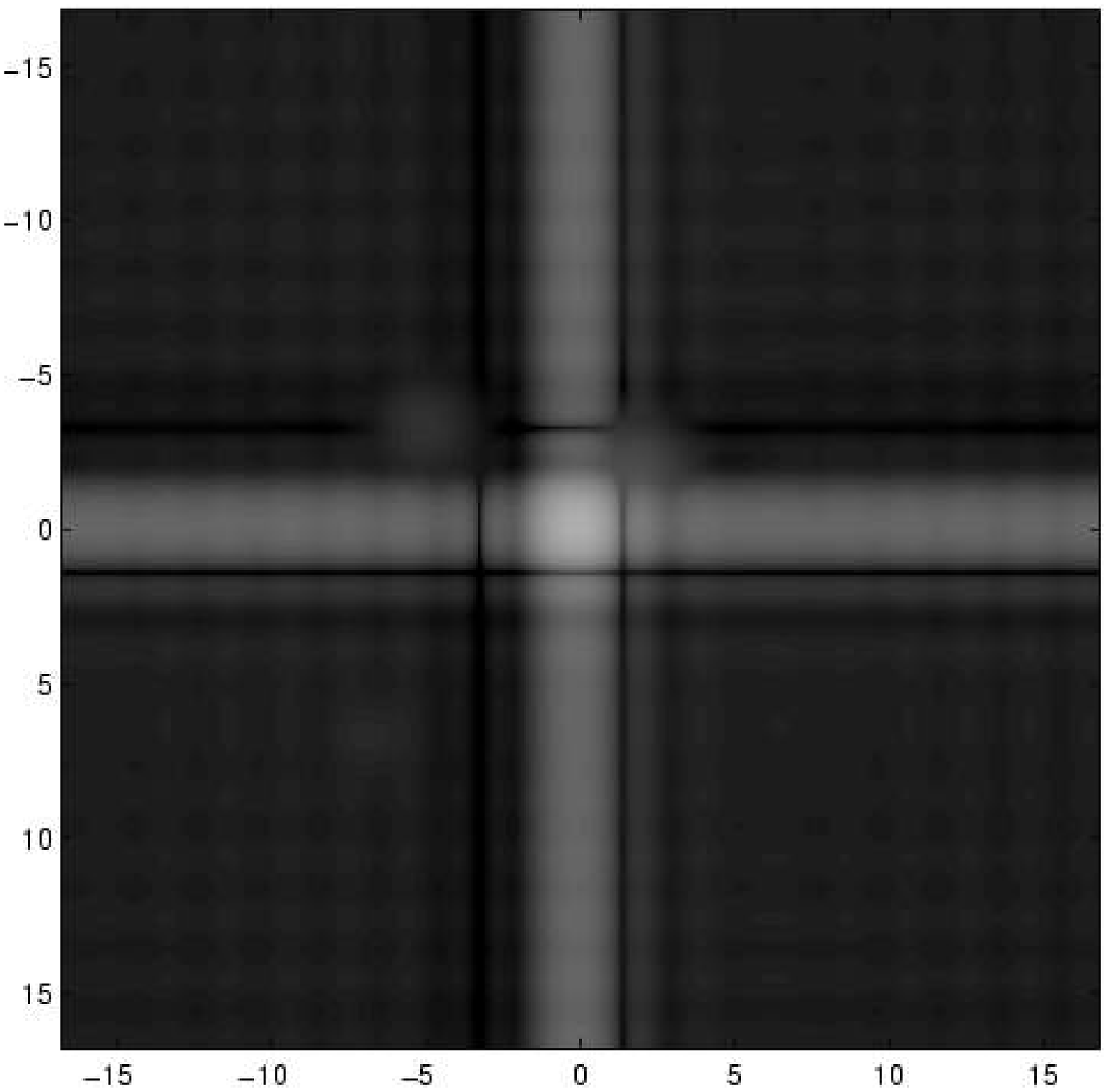}
\includegraphics[width=2.0in]{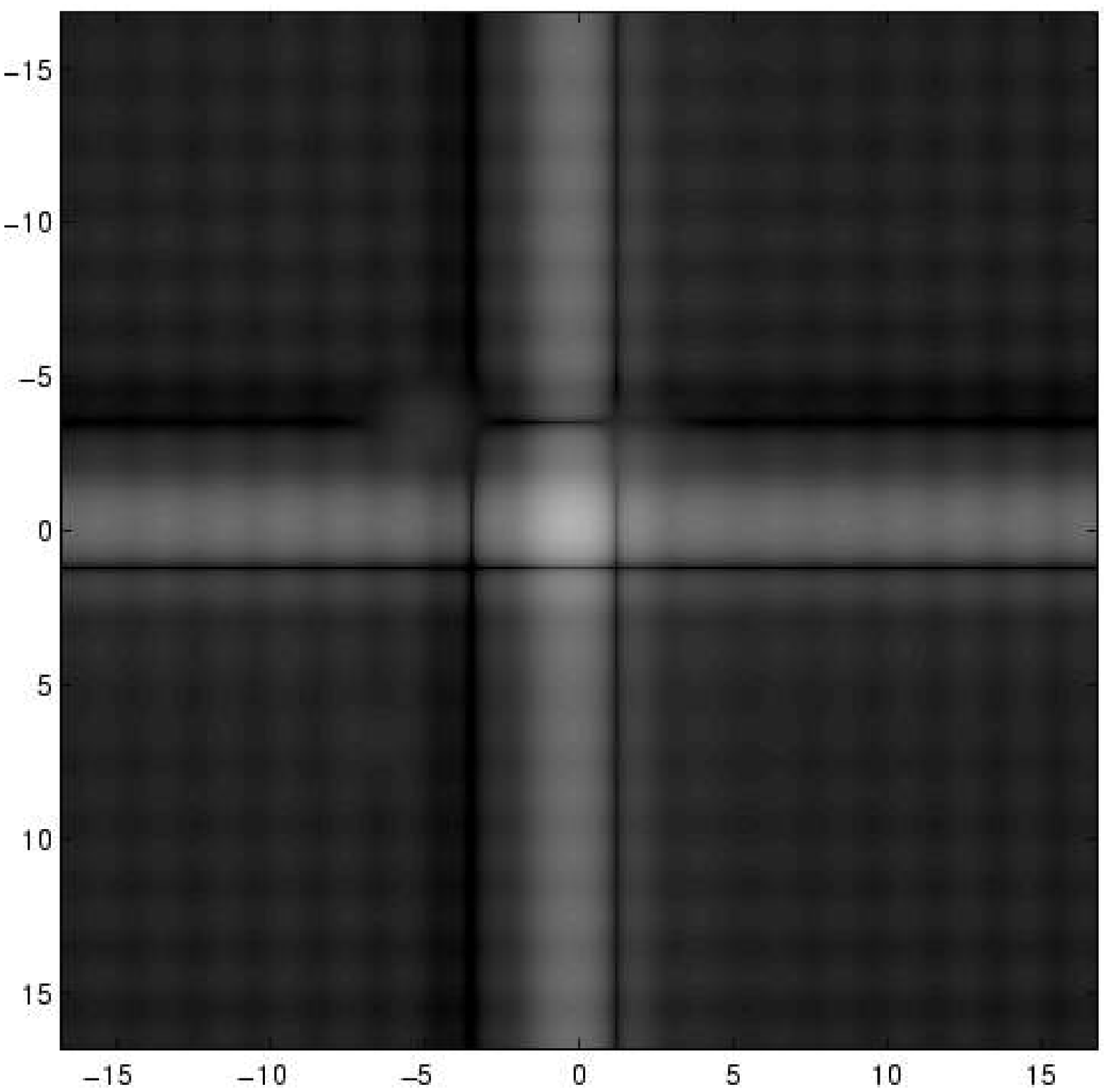}
\includegraphics[width=2.0in]{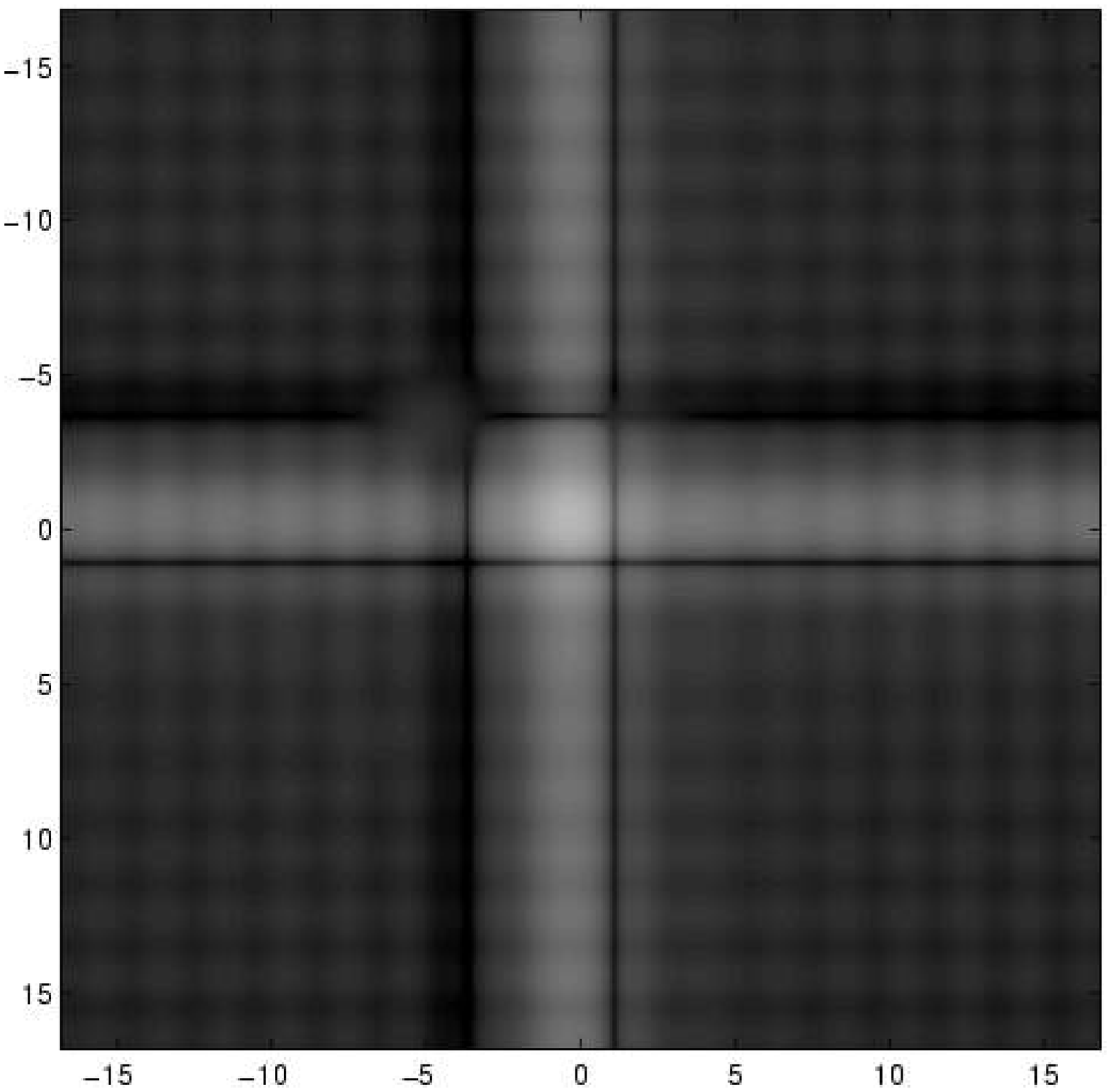}
\end{center}
\caption{
A simulated 3-planet star system as it would appear in a Lyot-style
coronagraph in which the entrance pupil is an open square with a plus-sign 
spider, the first image
plane has a ``plus'' shaped occulter of half-width $0.6 \lambda/D$, and a
rectangular mask in the Lyot plane.
This design includes a $2\%$ central obstruction.
{\em Top left.} A barcode mask designed to provide $10^{-5}$ contrast from 
$1.4 \lambda/D$ to $21 \lambda/D$ when used in conjunction with the
image-plane occulter.
{\em Top middle.} The corresponding rectangular mask.
The Airy throughput for this mask is $3.5\%$.  
{\em Top right.} The simulated image.  
The planet in the upper right quadrant is
located at $(1.7,1.7)$ and is $10^{-8}$ times as bright as the star.
The planet in the upper left quadrant is
located at $(-5.1,3.4)$ and is $10^{-9}$ times as bright as the star.
The planet in the lower left quadrant is
located at $(-6.8,-6.8)$ and is $3 \times 10^{-10}$ times as bright as the star.
The gray-scale image is logarithmically stretched to highlight the faint
planets. 
{\em Bottom.} The effect of pointing error.  From left to right, the pointing
error is $0.048$, $0.096$, and $0.144$.  The direction of the error is along
the diagonal into the first quadrant.
}
\label{fig4}
\end{figure}

The second example is the same as the first except that we impose
a central obstruction in the pupil-plane masks:
\begin{eqnarray} \label{16}
    A(x) = \left\{ \begin{array}{ll} 
     		1 & 0.01 \le |x| \le \half \\
		0 & \text{otherwise} 
           \end{array} \right. 
    \\[0.1in]
    \label{17}
    |L(x)| \quad \left\{ \begin{array}{ll} 
     		\le 1 & 0.01 \le |x| \le \half \\
		 =  0 & \text{otherwise} 
           \end{array} \right. 
\end{eqnarray}
The optimal barcode mask is shown in Figure \ref{fig4} together with
the corresponding rectangular mask.

\begin{figure}
\begin{center} 
\includegraphics[width=1.5in]{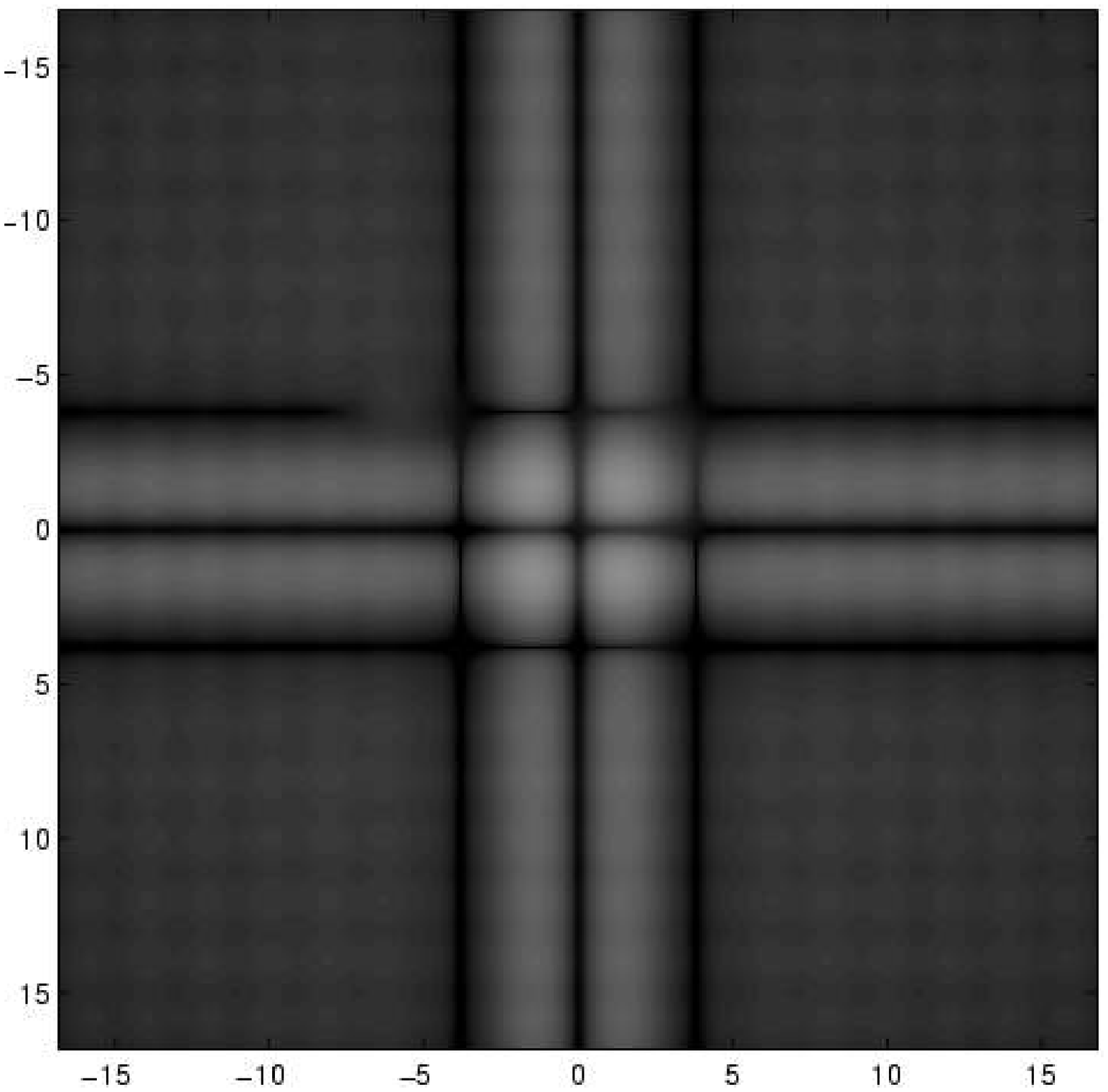}
\includegraphics[width=1.5in]{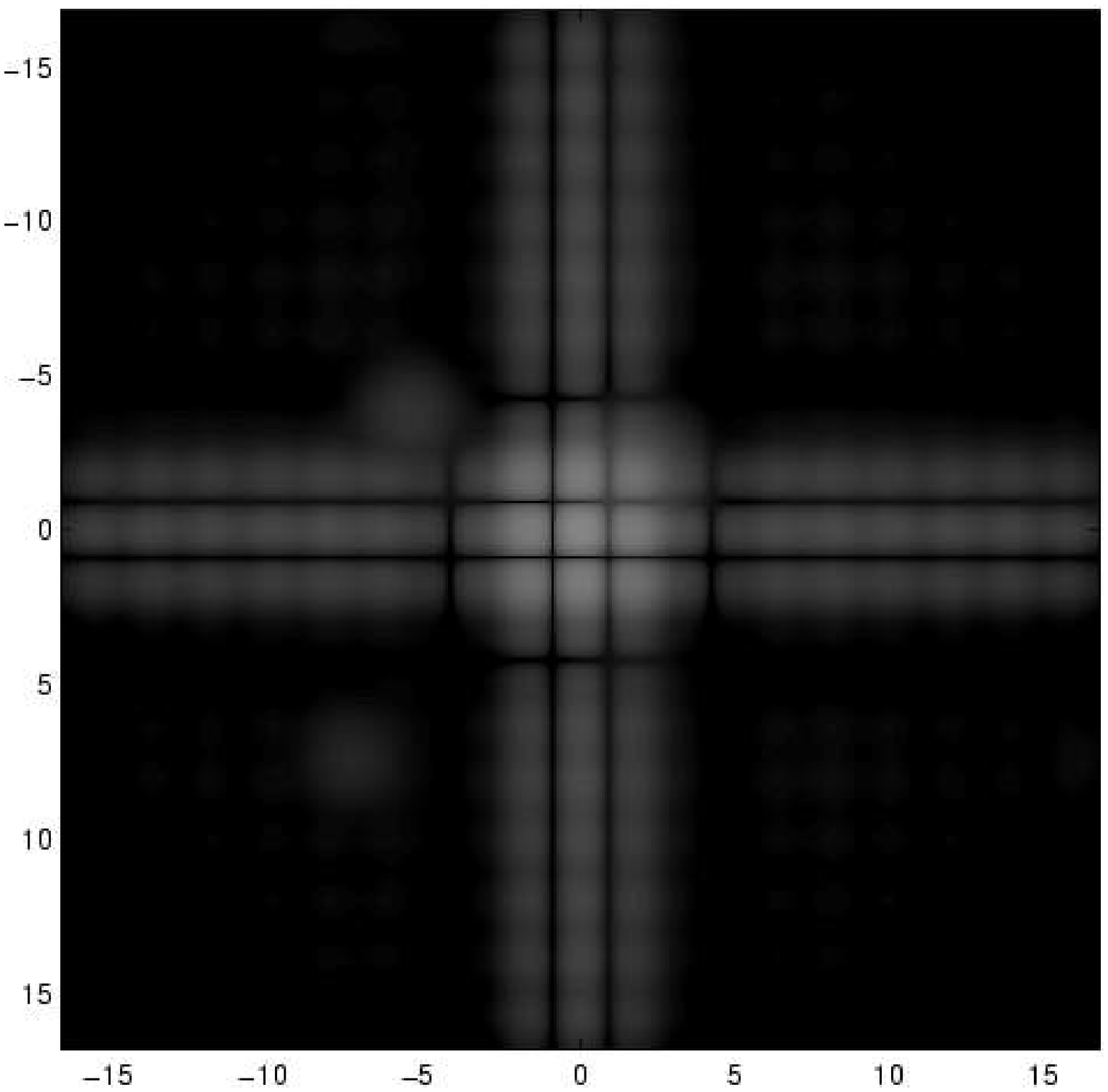}
\includegraphics[width=1.5in]{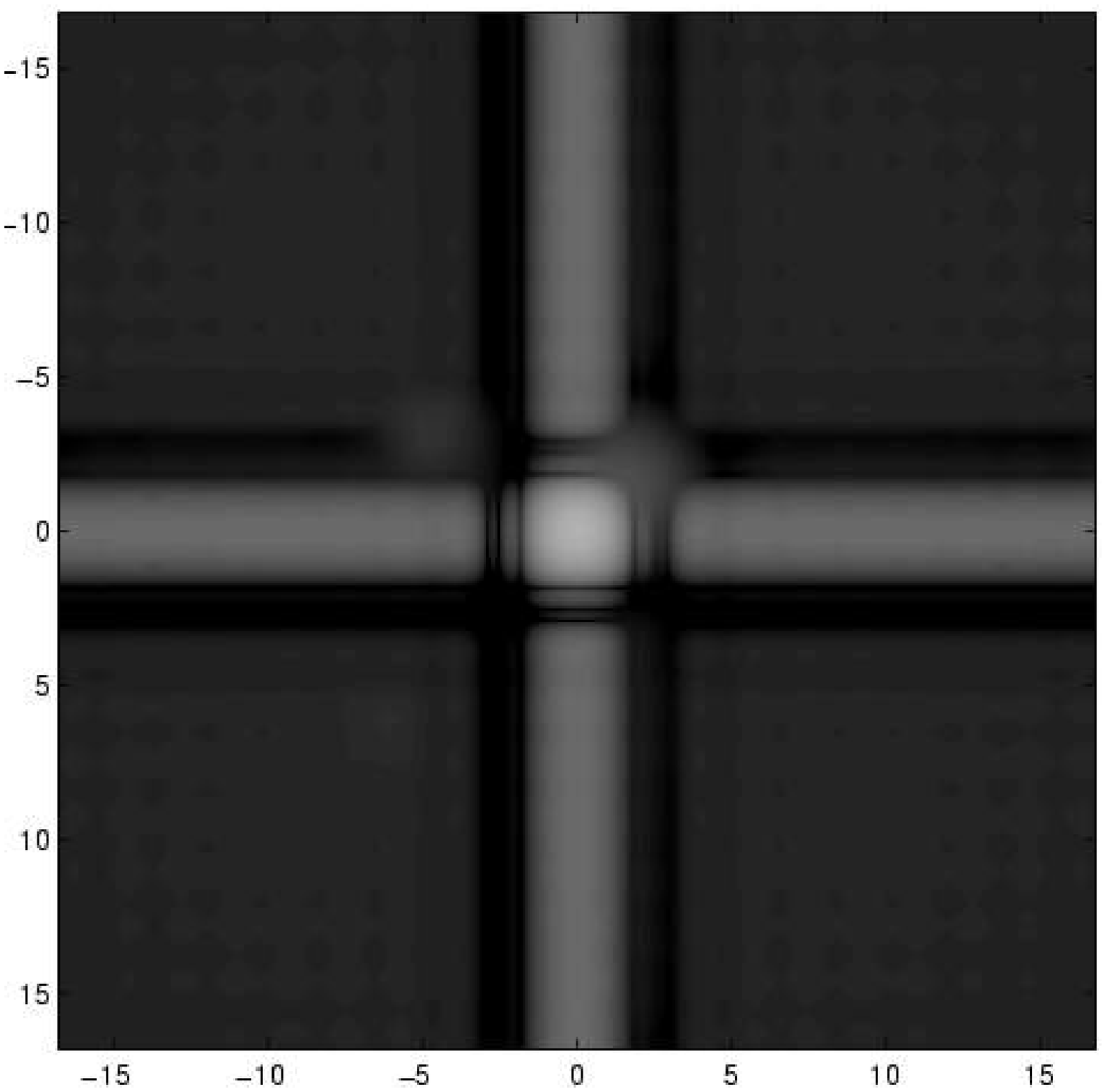}
\includegraphics[width=1.5in]{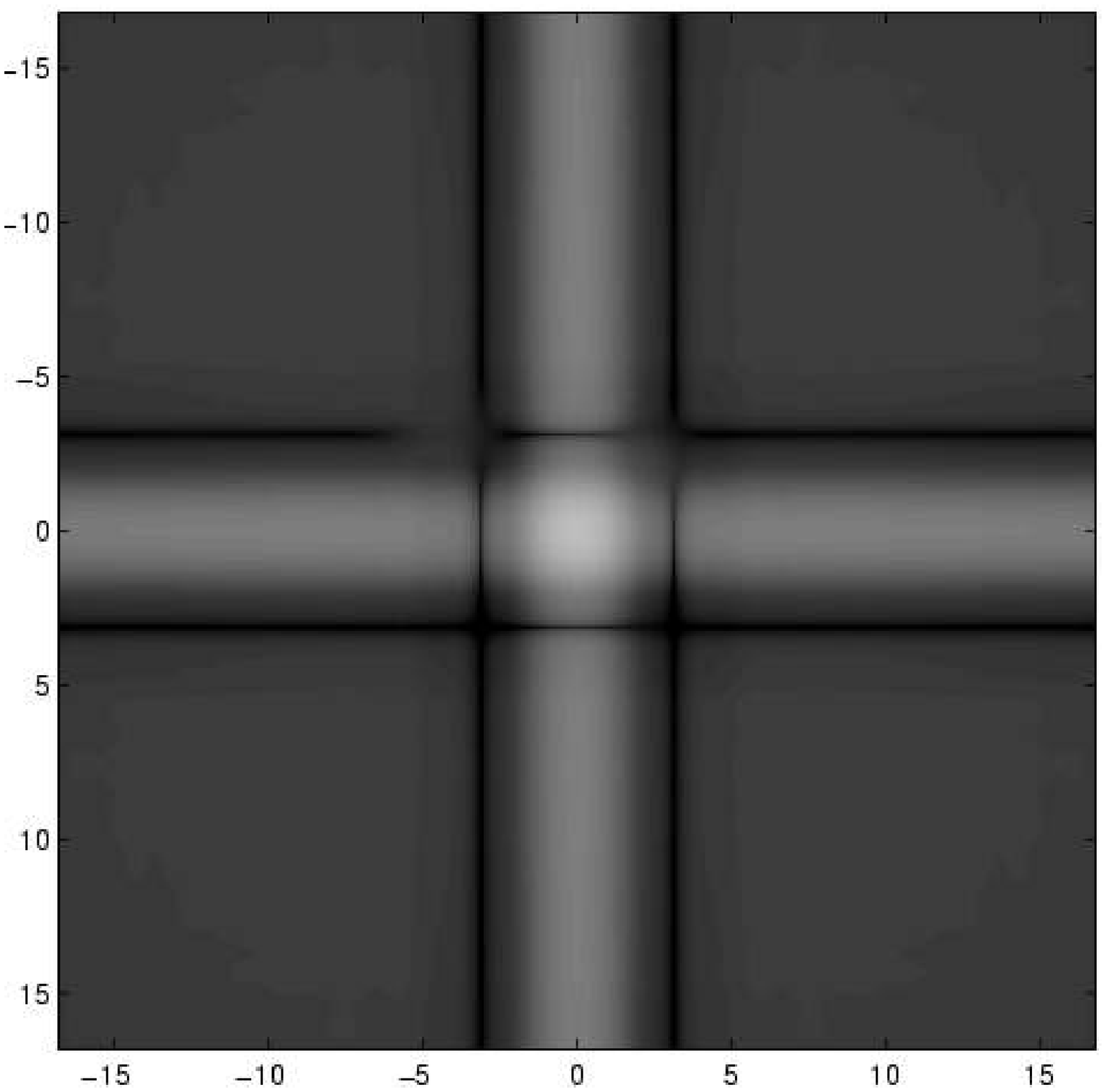}
\end{center}
\caption{
The simulated 3-planet star system as it would appear in the Lyot-style
coronagraph of Figure \ref{fig4} imaged at wavelengths shorter and longer than
the design point.  From left to right, the wavelengths are $85\%$, $90\%$,
$105\%$ and $110\%$ of the wavelength for which the system was designed.
}
\label{fig5}
\end{figure}

Since the designs presented in this section involve an image plane mask,
the associated PSFs depend on wavelength in a more complicated way than the 
designs given in Section \ref{sec:barcode}.  In fact, as Figure \ref{fig5}
shows, the loss of contrast occurs at wavelengths shorter than about $90\%$
and longer than $105\%$ of the design point.

\section{Conclusions}
\label{sec5}

In this paper we have presented some new masks for high-contrast imaging.
The best design 
\begin{enumerate}
\item achieves an inner working angle of $2.0 \lambda/D$ along
	the four diagonal directions,
\item includes a four-vane spider so that it can be used in an on-axis
	telescope design, and
\item consists of two orthogonally oriented barcode masks each one, by
	itself, being designed to provide only $10^{-5}$ level of contrast.
\end{enumerate}
We include an analysis of the effects of pointing error and chromaticity.

{\bf Acknowledgements.}
We are grateful for the continuing support we have received for this work
from NASA and JPL.
The first author also wishes to acknowledge support from the 
NSF (CCR-0098040) and the ONR (N00014-98-1-0036).

\bibliography{../lib/refs}   
\bibliographystyle{plainnat}   

\clearpage

\clearpage


\end{document}